# Software Architectures for Robotics Systems: A Systematic Mapping Study

## Aakash Ahmad[1], Muhammad Ali Babar[2, 1]

CREST - Centre for Research on Engineering Software Technologies

[1]Software and Systems Section, IT University of Copenhagen, Denmark

[2]School of Computer Science, University of Adelaide, Australia

[1]aahm@itu.dk, [2]ali.babar@adelaide.edu.au

## Abstract

**Context:** Software architecture related issues are important for robotic systems. Architecture-centric development and evolution of software for robotic systems has been attracting researchers' attention for more than two decades.

**Objective:** The objective of this work is to systematically *identify*, taxonomically *classify* and holistically *map* the existing solutions, research progress and trends that influence architecture-driven modeling, development and evolution of robotic software.

**Method:** We carried out a Systematic Mapping Study (SMS) to identify and analyze the relevant literature based on 56 peer-reviewed papers. We extract and synthesize the data from selected papers to (i) taxonomically classify the existing research and (ii) systematically map the solutions, frameworks, notations and evaluation methods to highlight the role of software architecture in robotic systems.

**Results and Conclusions:** We have identified eight distinct research themes that support architectural solutions to enable (i) *operations*, (ii) *evolution* and (iii) *development* specific activities of robotic software. The research in this area has progressed from *object-oriented* to *component-based* and now to *service-driven* robotics representing different architectural generations. The reported solutions have exploited model-driven, service-oriented and reverse engineering techniques since 2005. An emerging trend is *cloud robotics* that exploits the foundations of service-driven architectures to support an interconnected web of robots. The results of this SMS facilitate knowledge transfer – benefiting researchers and practitioners – focused on exploiting software architecture to model, develop and evolve robotic systems.

## 1. Introduction

Robotic systems are increasingly being integrated in various aspects of everyday life. The robotic applications range from mission critical [1] to infotainment and home service tasks [2, 3]. Robotic systems are expected to assist or replace their human counterparts for efficient and effective performance of all sorts of tasks such as industrial operations [6] or surgical procedures [4, 5]. A robotic system is a combination of various components – hardware (for system assembling) and software (for system operations) – that must be seamlessly integrated to enable a robotic system's function as expected. To support the vision of a robotic-driven world[1], academic research [1, 15, 21], industrial [10, 32] and open source solutions [9, 59] are striving to provide cost-effective and efficient solutions for the development, evolution and operations of robotics systems. Researchers [21] and practitioners [10] are increasingly focusing on exploiting software engineering methodologies to abstract complexities and enhance efficiency for modeling, developing, maintaining and evolving robotic systems cost-effectively [18, 53, 54].

---

[1] EUROP, the European Robotics Technology Platform is an industry-driven platform compromising the main stakeholders in robotics. EUROP aims at enabling research and practices through *Robotic Visions to 2020 and Beyond - The Strategic Research Agenda for Robotics in Europe*.

Software Architecture (SA) represents the global view of software systems by abstracting out the complexities of low-level design and implementation details [45]. SA plays a vital role in ensuring the fulfillment of functional and non-functional requirements [41]. It is considered that architecture-centric software development helps increase quality, modularity and reusability [21] through the use of patterns [47] and decrease complexity by applying model-driven development [46]. Researchers from different communities (such as robotics, software engineering, industrial engineering, and artificial intelligence) have exploited architectural models to design, reason about, and engineer robotic software. Architecture-centric robotics research and practice can be characterized by various trends such as: (i) object-oriented robotics (OO-R) enabling modularity [12], (ii) component-based robotics (CB-R) supporting reusability [13], and (iii) service-driven robotics (SD-R) exploiting dynamic composition [14] of software.

Since the early 90s, there has been a continuous stream of reported research on software architectures for robotic systems. It is a timely effort to analyze the collective impact of existing research on architectural solutions for robotic software. We decided to conduct a systematic mapping study by following the guidelines reported in [17] to investigate the state-of-the-art (published from *1991* to *2015*) that promotes architectural solutions to model, develop and evolve robotic software. The objective of this mapping study is to: *'systematically identify, analyse, and classify the software architectural solutions for robotic systems and provide a mapping of these solutions to highlight their potential, limitations along with emerging and future research trends'*. This study was motivated by a number of research questions, whose answers are expected to disseminate systematized knowledge among researchers and practitioners who are interested in the role of software architecture for robotic systems. The key contributions of this study are:

– Systematic selection and analysis of the collective impact of the existing research on software architecture related aspects of robotic systems for identifying (i) predominant *research themes*, (ii) architectural *solutions* for each of the themes, (iii) *framework* support along with (iii) *modeling notations*, *evaluation methodologies* and *application domain* of architectural solutions.

– Reflections on the (i) *progression* and maturation of research overtime, (ii) various *communities* and their contributions to research progression, along with (iii) existing and emerging *research trends*.

The results of this study suggest that **architectural solutions** support (i) *operations* (enabling information and resource sharing) (ii) *evolution* (with runtime adaptation and design-time re-engineering), along with (iii) development (such as modeling, designing and programming) of robotic software. A number of **architectural frameworks,** open source, academic and industrial solutions, are available for development;  UML or its derivative notations are predominantly used for **architectural modeling**. A majority of the architectural solutions support home service, mission critical and navigation robots. Architectural solutions in general have been evaluated using controlled experiments and simulation based techniques. However, there is a lack of reporting on **architecture specific evaluations** against functional and quality requirements [60]. An incremental progress of research from OO-R [12] and CB-R [13] and more recently SD-R [14] (**architectural generations**) have resulted in an emergence of cloud robotics [15];  model-driven robotics is also gaining momentum [46]. The results of this SMS benefit:

- Researchers who are interested in knowing the state-of-the-art of software architecture for robotics systems. The systematic classification of the existing research provides a body of knowledge for deriving new hypotheses to be tested and identifying the areas of future research.

- Practitioners who may be interested in understanding the reported solutions in terms of frameworks, modeling notations, tools and validation techniques for architectural development of robotic systems.

The rest of this paper is organized as follows. Section 2 briefly introduces robotic systems, software architecture and related studies. Section 3 describes the research methodology used. Classification and mapping of the research are reported in Section 4. Based on the classification, various architectural solutions and frameworks for robotic software are presented in Section 5. Modeling notations, validation techniques and application domains that complement architectural solutions are presented in Section 6. The frequency, sources and active communities highlight various demographic details of the published research in Section 7. We discuss the progression of research, future dimensions and the emergence of various architectural trends in Section 8. The validity threats are presented in Section 9. Section 10 presents the key conclusions from this study. Appendix A lists the selected literature for mapping study. Appendix B presents extended details of the research methodology.

# 2. Background and Related Studies

In this section, we briefly introduce robotic systems (in Section 2.1) and software architecture (Section 2.2). We also discuss some existing secondary studies (Section 2.3) that are related to our SMS.

## 2.1 An Overview of Robotic Software Systems

A robotic system is a combination of hardware and software components (two distinct layers) that can be integrated to build a robot [10, 11]. The ISO 8373:2012 standard provides a vocabulary of robots and various robotic devices that operate in industrial and non-industrial environments. The hardware components such as the sensors, robotic arms, navigation panels enable the assembly of a robot. Hardware components are controlled and manipulated by Control Layer that is essentially a collection of drivers (as system specific code) to interact with the hardware as in Figure 1. For more complex functions of a robot, specialized software is provided for integration and coordination of hardware components to manipulate the robotic behavior. In Figure 1, this refers to as Application Layer that utilizes the control layer to support robotic operations. For example, considering a home service robot [7], the control layer provides a driver that enables access to a robotic arm. Depending on specific requirements, a software system must be provided. Such software system is expected to utilize drivers (from control layer) to enable the movement of arm for home service robot at certain degrees of precision and/or providing logic to avoid any obstacles.

Figure 1 highlights that from a software perspective robotic systems are an example of a layered design, where each layer encapsulates a specific functionality and depends on the layer(s) above or below to complete a system's functionality. The focus of this study is to review the research state-of-the-art for architectural solutions supporting robotic software – *application layer*.

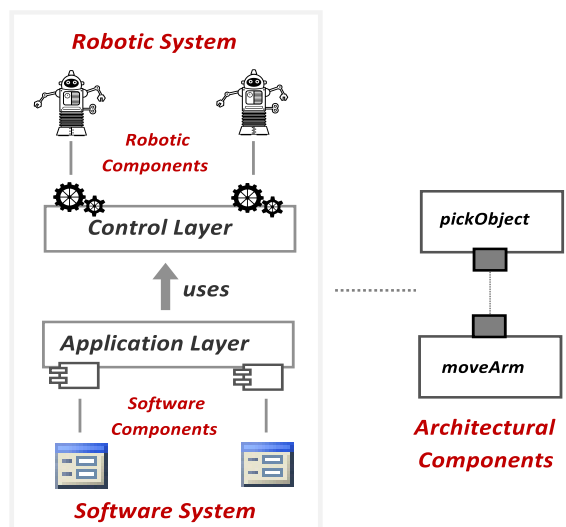

Figure 1. A Reference Model for Robotic Software Systems

## 2.2 Software Architecture

The ISO/IEC/IEEE 42010 is a standard for architecture description of software systems that represents architecture as an aggregated or high-level view of software in terms of architectural components (computational elements) and connectors (interconnections between components). This is also referred to as the component and connector (C&C) view of a system; architectural components as computational elements represent an abstraction of executable code [41] and communicate to each other using connectors. For example, considering Figure 1, the classes or modules of code that supports movement of robotic arm can be packaged into *moveArm* component that communicates with *pickObject* component. With C&C view the complex and implementation specific details (i.e., modules and function calls) are abstracted to highlight that picking or moving an object depends on the movement of a robotic arm.

Architecture models are beneficial for stakeholders in general and software designer and architects in particular to model and evolve software systems. An architectural model helps focus on higher level of abstractions, when implementation details have yet to be decided or are not considered relevant during reasoning about high level software and system decisions. For example, model-driven engineering [46] allows requirements of a domain (e.g., movement of robotic arm) to be mapped to architecture models or components (e.g., *moveArm*) to generate executable code in a stepwise manner [13, 21].

## 2.3 Existing Secondary Studies on Software for Robotic Systems

We found two types of secondary studies related to our study, as in Table 1. We briefly discuss these studies in terms of their scope and contributions to justify the needs for our study. Table 1 presents a summary of these studies in terms of study reference, focus, publication year with individual details below.

### 2.3.1 Systematic Literature Reviews of Software Engineering for Robotics

*Oliviera et al*. [18] have reported a systematic review of Service-Oriented Development of Robotic Systems based on 39 primary studies published from 2005 to 2011. Their review reports the solutions that support design, development and operation of robotic systems based on software services using service oriented approaches. Their review highlights the needs for improving reuse, productivity and quality of service-oriented robotic systems - providing a catalogue of solutions to develop service-driven robotics.

*Pons et al*. [19] have reported another systematic review of Software Engineering Approaches for Robotics based on 67 studies that were published from 1999 to 2011. They highlight the prominent trends of software engineering techniques for robotic software. The review highlights the application of (i) component based, (ii) service oriented as well as (iii) model driven development of robotics as the emerging research trends.

Table 1. A Summary of secondary studies on Software Engineering for Robotics.

| Study Reference | Study Focus | Publication Year | Number of Studies |
|---|---|---|---|
| *Systematic Literature Reviews* | | | |
| Oliviera et al. [18] | *Service-oriented Robotics* | 2013 | 39 |
| Pons et al. [19] | *Software Engineering for Robotics* | 2012 | 67 |
| *Survey-based Studies* | | | |
| Zhi et al. [53] | *Robotic Coordination Systems* | 2013 | N/A |
| Elkady et al. [54] | *Robotic Middleware* | 2012 | 21 |
| Kramer et al. [55] | *Robotic Development Environments* | 2007 | 09 |

### 2.3.2 Survey-based Studies of Robotic Software

*Zhi et al*. [53] have reported a survey-based study of Multi-Robot Coordination Systems. As shown in Table 1, their study is aimed at identifying research challenges and problems including communication mechanisms, planning strategies and decision-making structures for robotics. These problems are discussed in the context of cooperative and competitive environments in which team of robots coordinate to accomplish specific missions. A systematic classification and comparison of the problems helps to identify the dimensions of future research such as energy efficiency, decision making, and heterogeneous mobile robots.

*Elkady et al.* [54] have reported a survey study of Robotics Middleware by reviewing over twenty solutions such as OPRoS, MRDS, CLARAty. The survey evaluates the strengths and limitations of robotic middleware and provides a set of guidelines to select the most appropriate middleware based on the requirements of robotic systems. The study highlights some important attributes of each middleware such as simulation environment, support for distributed environment, fault detection and recovery and behavior coordination capabilities to help developers while selecting an appropriate middleware for robotics software.

*Kramer et al.* [55] have reported another survey study of Robotic Development Environments (RDEs) to highlight the importance of development environments and frameworks for robotic systems or specifically mobile robots. The survey primarily aims at analyzing the usability and impact of the major RDEs (Player, Miro, and MARIE). The survey provides guidelines to analyze the strength and limitations that help researchers/practitioners to select an appropriate RDE. A systematic comparison and evaluation of these RDEs also highlights areas of future research on the development and application of RDEs.

The surveys reported in [54, 55] suggest the terms *middleware* and *development environments* are complementary and often interchangeable. Elkady and Sobh [54] refer to frameworks like Player, Pyro, and Miro as middleware to abstract platform and hardware specific details. On the other hand, Karmer and Scheutz [55] describe the above-mentioned frameworks as open source RDEs for architectural development of mobile robots. It is vital to mention about [61] that presents the state-of-the-art in *robot programming systems* with a distinction between manual and automatic programming. The survey suggests that software architectures are important to both of these programming systems, as architecture provides underlying support to model and develop robotic software. We assert that this study complements the existing body of research on architecture-centric development of robotic software. Our study specifically addresses various architectural aspects, challenges, and their solutions for robotic systems by investigating the state-of-the-art that has progressed over more than two decades (1991 to 2015).

# 3. Research Methodology

We used Systematic Mapping Study (SMS) method [17] that involves a three step process (in Figure 2): (i) planning a study, (ii) data collection and synthesis, and (iii) mapping and documenting results. A systematic approach for a review reduces bias in identifying, selecting, synthesizing the data and reporting results.

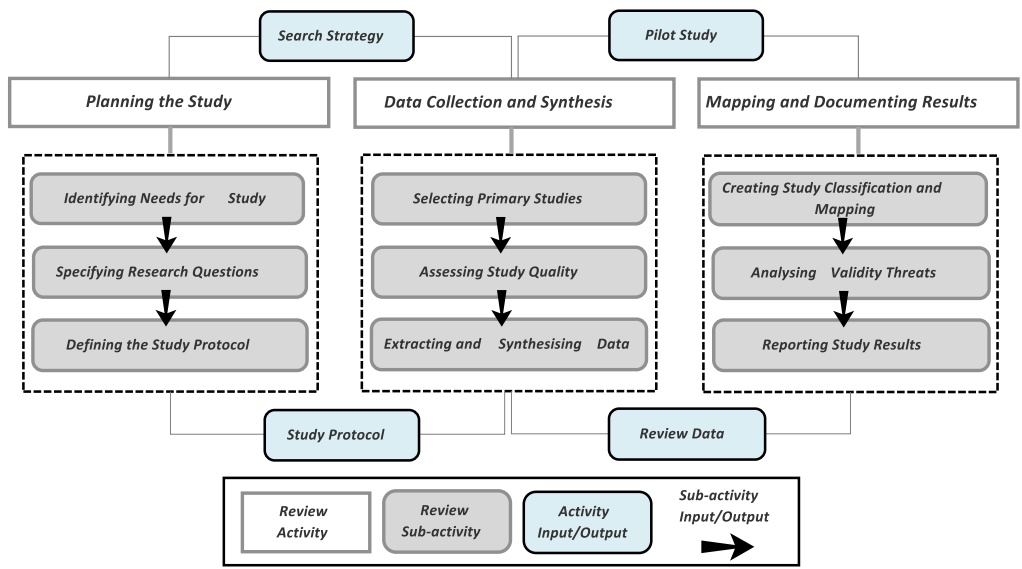

Figure 2. An Overview of the Methodology for Mapping Study

## 3.1 Planning the Study

A study plan purports to identify and justify the needs of an SMS and to define the scope of the study with specific research questions. The study plan helps formulate a protocol and create the search strategy that guides the rest of the steps of the mapping study as illustrated in Figure 2.

### 3.1.1 Identify the Needs for Mapping Study

Despite a multi-disciplinary and continuously growing research for more than twenty years, there was no effort to systematically select, analyze, and report the peer-reviewed research on the progression, maturation and emerging trends of architectural solutions for robotic software. In contrast to [18, 19], that are focused on service-orientation or in general software solutions for robotics; our study is focused on generic architectural solutions for robotic software. Compared to [18], our study is aimed at going beyond SOA and provides a more comprehensive mapping and review of other architectural solutions. Before conducting the SMS, we must ensure that a similar study to our review has not been conducted or published. Therefore, we also searched the IEEE Xplore, ACM Digital Library, Springer Link and Science Direct (on 15/08/2014) with the following search string in Listing 1 to identify the relevant secondary studies. Specifically, Listing 1 presents the search string to identify any relevant (survey/review-based) secondary studies on software architecture for robotic systems. Based on the literature identification with the following search string, none of the publications that we retrieved (cf. Table 1) were aimed at answering the outlined research questions below that had motivated our mapping study.

> ("*Systematic Literature Review*" **OR** "*Systematic Mapping*" **OR** "*Study*" **OR** "*Survey*")
> **AND**
> ("*Software Architecture*" **OR** "*Software Component*" **OR** "*Software Framework*" **OR** "*Software Engineering*")
> **AND**
> ("*Robot*" **OR** "*Robotic*" **OR** "*Humanoid*")

Listing 1. Search String to Identify the Relevant Secondary Studies.

### 3.1.2 Specifying the Research Questions

We formulated three research questions to be answered by our study. Each of the main research questions has sub-questions for fine-grained investigation and presentation of the results.

**RQ 1: What is the state-of-research on software architecture-based solutions for robotic systems?**
The objectives of this question can be met by answering the following sub-questions.

– *RQ 1.1: What research themes have been identified and how they can be classified?*
Objective: The identification of research themes is fundamental to a systematic analysis and a collective impact of the existing research on the topics to be investigated. We aim to derive a classification scheme (i.e., taxonomy, systematic map [45] or classification framework [44]) as a foundation to identify, taxonomically classify various types of (distinct or overlapping) themes reported in the selected studies.

– *RQ 1.2: What types of architectural solutions have been reported for robotic software?*
Objective: Based on the classification, we intend to identify the research challenges and architectural solutions that could address these challenges for robotic software. The classified themes (RQ 1.1) could allow us to abstract the details of individual solutions and analyze a collective impact (strengths and limitations) of the existing research.

– *RQ 1.3: What architectural frameworks have been provided to support the solutions?*
Objective: As per ISO/IEC/IEEE 42010, an architectural framework supports various architecting activities (such as modeling, analysis, and development) to facilitate architecture-centric solutions [54, 55]. Therefore, we investigate framework support for architectural solutions (RQ 1.2) that could avoid re-invention and promote reuse of existing tools and techniques for developing complex solutions.

– *RQ 1.4: What architectural notations have been exploited for solution representation?*

*Objective*: In order to develop any architectural solution, the solution modeling or representation is a fundamental step. The identification of various architectural notations could help with analyzing, what are the prominent notations and how are they exploited in solution modeling [45, 47] – also complementing RQ 1.2.

– *RQ 1.5: What validation methods have been employed to evaluate the solutions?*

*Objective*: The effectiveness of a solution is analyzed based on its validation [20, 37]. A study of the used validation methods allows us to analyze the types of evaluation methods and techniques used. Moreover, the presentation of results can highlight the findings about overall solution validations [37] or architecture-specific evaluations [60] for robotic software.

– *RQ 1.6: What were the application domains for architectural solutions?*

Objective: Robotic solutions are applied to various domains. Mission critical [1], infotainment and home service [2, 3], medical robotic [4] are some of the domains where robotics systems are being used. An answer to this question would highlight the specific domains where architectural solutions are pre-dominant.

**RQ 2 – What are the demographic details of research in terms of publication years, sources and active communities?**

To answer different aspects of research demography, we decompose the questions as follows.

– *RQ 2.1: What is the publication frequency and fora of research over the years?*

Objective: To highlight the various aspects including the year and frequency of research publications, publication focus and fora allows us to reflect on a temporal progression of the research [45, 47].

– *RQ 2.2: What are the prominent venues of publication and the types of published research?*

*Objective*: To identify the prominent publication venues (i.e., publication sources) in terms of journals and conference series along with the type of research that is published through those venues.

– *RQ 2.3: What research communities are active on software architectural solutions for robotic systems?*

*Objective*: To identify the extent to which the research on software architectural solutions for robotic systems can be regarded as multi-disciplinary. We aim to identify this by highlighting the communities (i.e., software engineering and robotics) actively focused on researching software architecture-driven robotics.

**RQ 3 - What were the past trends and what types of existing and emerging trends could be identified for architecture-based solutions for robotic systems?**

To discuss the past, present and possible future research trends, we answer the following questions.

– *RQ 3.1: What are the past trends of research on architecting robotic software?*

*Objective*:  To identify the research trends that emerged in the past that can help us to discuss how they contributed to the development of existing solutions or active trends.

– *RQ 3.2: What are the emerging trends of research on software architecture for robotics?*

*Objective*: Based on analyzing the past and current trends, we could present the emerging trends that may represent emerging problems and innovative solutions that address them as dimensions of future research.

### 3.1.3 Defining the Protocol for SMS

The study protocol includes (i) a study's scope based on research questions to be answered (cf. Section 3.1.2), (ii) search string and search strategies (based on RQs) to identify, include/exclude and qualitatively analyze the relevant literature. A study's protocol documents methodological details and logistical procedures for a research study. The protocol also allows us to evaluate different activities and their outcomes. We developed a study protocol that was internally reviewed a few times for improvement. We also have our study's protocol externally evaluated for refinements and reducing the bias. We performed a pilot study by reviewing 15 (more than 25%) of the included studies. The pilot study purported to reduce the study identification bias and to refine the process for (i) identification of primary studies, (ii) extraction of data from these studies and (iii) synthesizing the results. Based on the protocol review and pilot study, we expanded the review scope in the context of relevant studies ([18, 19], cf. Table 1), improved search strategies and refined the study inclusion/exclusion and qualitative assessment criteria [50]. We also evaluated the study protocol with details provided later in this section.

## 3.2 Searching the Relevant Literature – Primary Studies

After defining the protocol, we followed the steps to collect and synthesize the data (from Figure 2). In Appendix B, we discuss the steps taken for (i) *selection of primary studies* (**B.1, Table I**), (ii) *screening and assessment of the studies* (**B.2, Table I and Table II**), and (iii) *data extraction for synthesis* (**B.3, Table IV**).

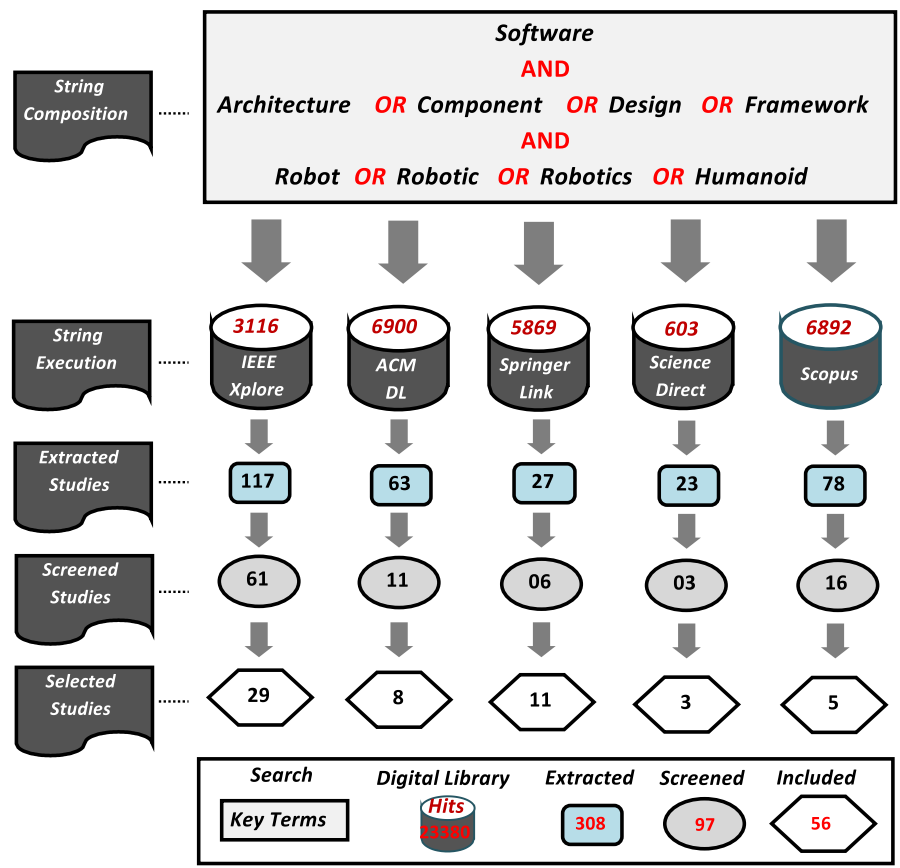

Figure 3. Summary of the Literature Search Process with Search String

Figure 3 shows the search process used for this study as shown in Figure 3. The above-mentioned research questions helped us to identify a set of keywords that were used to build a search string that was applied to five databases shown in Figure 3. We limited our search to the peer-reviewed literature from years 1991 to 2015 (15/08/2015). The year 1991 was chosen as the initial search found no earlier results related to any of the research questions with 23380 hits. In the primary search process (**Appendix B, Section B.1**), we focused on title and abstract, therefore, it resulted in a high number of studies that were not relevant, which we refined with secondary search process - limiting the extracted studies to 308 in total. In order to identify and select primary studies, the search string was customized as per individual databases for effective search [16] (detailed in **Appendix B**). Please note that the first search string (cf. Section 3.1.1) helped us to identify the relevant secondary studies - systematic reviews and surveys in Table 1 - on software architecture for robotic systems. In contrast, the search string in Figure 3 aims to identify the primary studies listed in Appendix A, methods and techniques, for architecture-driven robotics software. Based on screening and qualitative assessment of the extracted studies (**Appendix B, Section B.2**), out of 97 a total of 56 studies were selected for inclusion in this study presented in **Appendix A**.

### 3.2.1 Evaluating the Protocol for SMS

Once the protocol is defined, the guidelines for conducting the SLR and SMS [16, 17, 50] suggest the needs to internally and externally evaluate the study protocol before its execution. We performed both the internal and external evaluation of the protocol to eliminate or minimize the possible bias. We focused on specifically evaluating steps that included (i) *identification and qualitative assessment of the primary studies*, (ii) *consistency of data extraction and reporting*, and finally (iii) *data synthesis and results reporting*.

***Internal Evaluation*** - As a team of two researchers, we focused on a structured representation of the information (e.g.; Qualitative assessment of studies – Table III, Data extraction template – Table IV in **Appendix B**) for an objective interpretation and evaluation of the methodology steps. First of all, both the researchers executed the search string on the selected digital libraries (cf. Figure 3) individually and then shared the search results. We also conducted a pilot search first to refine the search strings. For example, the terms *software architecture* and *software design* are complementary and virtually synonymous. We also included the term software design in the search string that led to a significant increase in the number of identified studies and the efforts to retrieve the most relevant. We excluded the terms like 'Software Structure' or 'Software Styles' in order to minimize the irrelevant literature. Once the consistency of search results was ensured, the first researcher identified the relevant studies, maintaining their references and derived a multi-criteria assessment of the study quality (cf. Table IV, **Appendix B**). The second researcher cross-checked the results of literature identification (randomly selecting and checking results with 2/5 of the digital libraries) and assessing studies against the qualitative assessment checklist.

***External Evaluation*** - As a two-step process was performed by a researcher external to our team, who had expertise in conducting SLRs and whose research interests lied in the area of software architecture. In the first step, we shared the RQs, identified studies and the data extraction form. Based on his feedback and recommendations, we refined the research questions and made necessary adjustments to the data extraction template (cf. Table IV, **Appendix B**). For example, a suggestion was to distinguish between the presentation of solution validation and architecture evaluations (attributes M07, M08 in Table IV – Data Extraction Form). In the second phase, due to time constraints instead of sharing the detailed results, we only shared the data extraction form and research questions. Based on the external feedback, we refined and finalized the data extraction form before capturing data and synthesizing results. Some possible threats to the validity of research are detailed later after discussion of the results. Based on the extracted data (**Appendix B, Section B.3**) and the objectives of the RQs (Section 3.1), we present the results of SMS in Section 4 to Section 8.

# 4. A Taxonomical Classification and Mapping of the Research

First of all, to analyze the state-of-research (RQ 1) we answer a number of questions (RQ 1.1 – RQ 1.6) in subsequent sections. In this section, we specifically focus on the results for **RQ 1.1** that answers: *What research themes have been identified and how they can be classified?* To analyze a collective impact of existing research, first we identified the predominant research themes by applying thematic analysis [46] on individual studies and then classified and presented them as taxonomy in Figure 4. The taxonomy in Figure 4 classifies the identified themes as an overview of the existing research (Section 4.1) and guides the discussion on the results for this mapping study (Section 4.2).

## 4.1 A Taxonomy of the Research Themes

The taxonomy in Figure 4 provides a systematic *identification*, *naming* and *classification* of various research themes based on the similarity or distinctions of their relative contributions for a systematic mapping. By analyzing some relevant studies (e.g., [44, 38]) and following some of the guidelines from *ACM Computing Classification System*[2] and *Computing Research Repository*[3] we derived the following categories:

---

1. **Generic Classification** that highlights the role of software architecture to support the *operations*, *evolution* (post-deployment) and *development* (pre-deployment) phases of robotic systems. The generic classification is used to organize the results (reviewed studies) into three distinct areas. Specifically, in the context of Figure 4 the literature is generally classified into approaches for i) *operational* ii) *evolution* specific and iii) *development* related issues of robotics.

2. **Thematic Classification** extends the generic classification by adding details based on the primary focus of research in a collection of related studies to identify and represent the recurring research themes using thematic analysis [46]. We identified three predominant themes as i) *coordination* (11 studies, i.e., 20% approx. of the reviewed studies), ii) *adaptation and reengineering* (13 studies, i.e., 23%), and iii) *modeling, design and programming* (32 studies, i.e., 57%) of robotics in Figure 4.

3. **Sub-thematic Classification** provides a fine-grained refinement of above-mentioned three themes with eight distinct sub-themes. Specifically, the research on architectural support to enable robotic coordination can be classified into two distinct themes (i) *information fusion* (05 studies, i.e., 09% approx. of total reviewed studies) and *distributed resource access* (06 studies, i.e., 11% approx.). Robotic adaptation and reengineering solutions are decomposed into three types namely (i) fault tolerance (03 studies, i.e., 05%), ii) reconfiguration (09 studies, i.e., 16%), and iii) reengineering (01 studies, i.e., 02%). The studies on robotic development are sub-classified to support architecture-driven (i) modeling (07 studies, i.e., 12.5%), (ii) design (18 studies, i.e., 32%), and (iii) programming (07 studies, i.e., 12.5%) of robotic software. For example, in the classification scheme; the generic classification called *Evolution* (E) has a specific research theme named *Adaptation and Re-engineering* that has three sub-themes *Fault Tolerance* [S4, S8], *Reconfiguration* [S5, S7, S23][4] and *Reengineering* [S14]. In contrast, supporting the fusion of information (a sub-theme) is an operational challenge (O) for mission critical robots that require coordination (a research theme) between a team of robots for collaborative completion of a task such as search and rescues mission [S9].

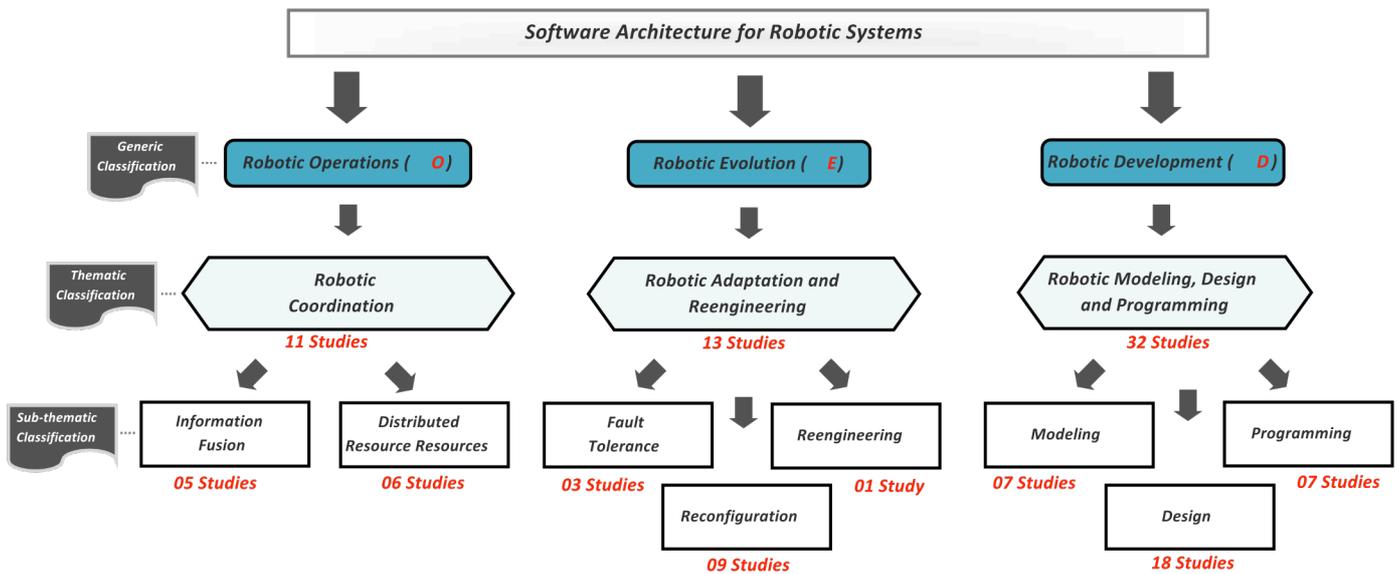

Figure 4. A Taxonomy of Research Themes on Software Architecture for Robotics.

Some studies could be classified into more than one theme, provided that their contributions were relevant to multiple themes. For example the study [S39] has been classified as a solution to robotic reconfiguration (i.e., evolution) by exploiting model-driven adaptation. Since, models are required to support reconfiguration it could also be classified under robotic design (model-driven robotic development [S40]) category.

---

[3] Computing Research Repository (CoRR): http://arxiv.org/corr/home

[4] The notation [Sn] (n is a number) represents a reference to studies included in the review, which are listed in the **Appendix A**. The notation also maintains a distinction between the bibliography and selected literature for this study.

Based on the discussion and consensus among the researchers, the study [S39] has been only classified under sub-theme reconfiguration – as the primary contribution is not to establish model-driven techniques but to exploit them to address the challenges of robotic reconfiguration. Architecture-driven reengineering of robotic was also identified in only one study [S14], it was decided to classify it under robotic adaptation and reengineering that supports design time evolution of a legacy robotic system.

## 4.2 A Mapping of Research Themes and Corresponding Evidence

While the taxonomy (cf. Figure 4) provided a broader classification of existing research, synthesizing and interpreting the results suggested the needs for a mapping of the identified themes, sub-themes and their corresponding evidence (relevant studies) in Figure 5 – a three dimensional mapping detailed below.

1. Three distinct research themes (adopted from Figure 4 – Thematic Classification) on x-axis (right). We refer to this as *thematic facet* of mapping in Figure 5.

2. Yearly distribution of relevant studies (as temporal distribution of evidence) on x-axis (left). We refer to this as *temporal distribution facet* of published evidence in Figure 5.

3. We map eight distinct sub-themes (from Figure 4 – Sub-themes) that present the specific contributions of research on y-axis. We refer to this as *contribution facet* of mapping in Figure 5.

The white rectangles on x-axis has two primary purposes: (i) the right side highlights the relevant studies as the identified evidence corresponding to individual themes and their sub-themes, ii) the left side presents the temporal distribution or publication frequency of the evidence. The interpretation of the thematic mappings in Figure 5 is based on locating a given research theme (x-axis) that allows the identification and mapping of its sub-themes (y-axis) along with identification of each relevant study in the circle. For example, the representation of [S5] highlights that '*a specific contribution of this research (*Publication Year 2009*) is to support architecture-driven reconfiguration of robotic systems as the identified evidence about the role of software architecture to support dynamic adaptation (i.e, runtime evolution) of robotic systems*'.

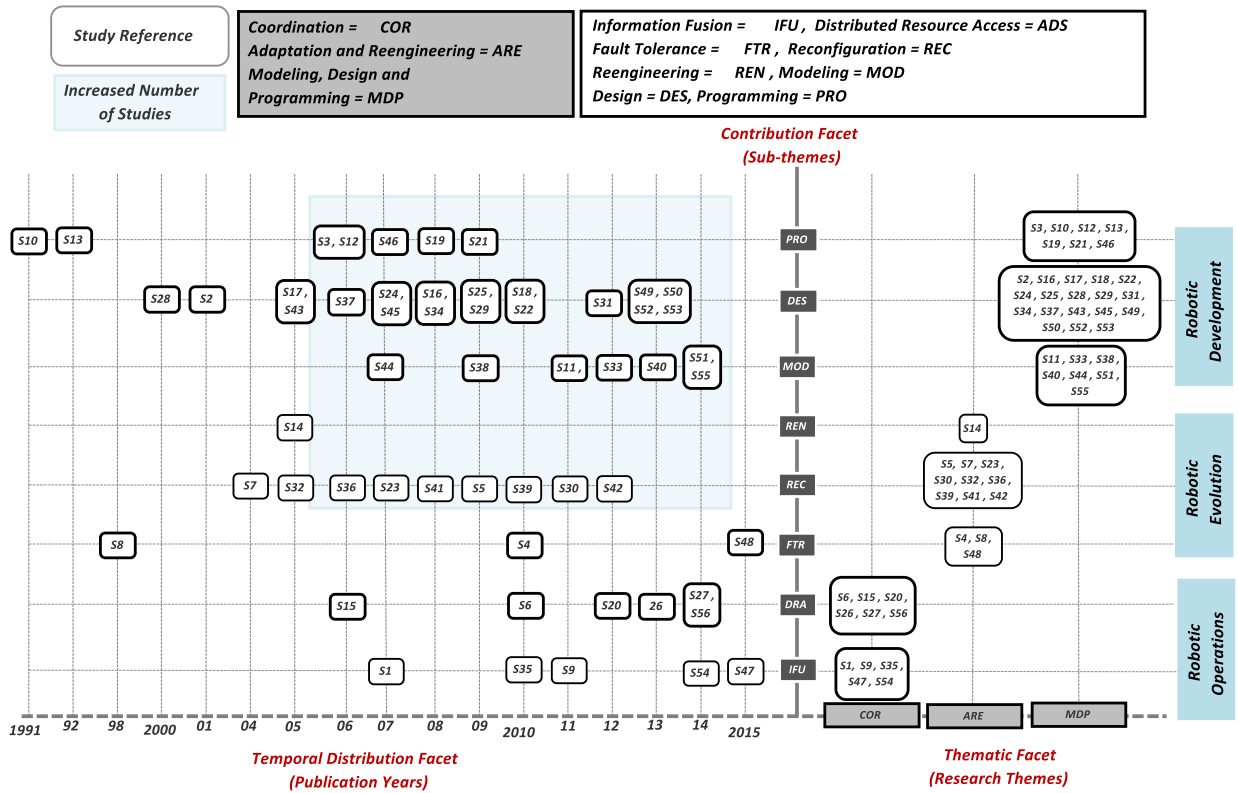

Figure 5. A Mapping of Research Themes, Sub-themes and Yearly Distribution of Relevant Studies.

***Interpretations and Usage of Classification and Mapping Schemes*** The mapping diagram in Figure 5 have diverse and multiple interpretations based on the intent of the analysis. For example, The yearly distribution highlights that '*in order to support evolution; reconfiguration of robotics is the most researched* *area among others with a total of 9 studies published between 2004 to 2012, while reengineering got little attention with only 1 study in 2005'.* In this context, following two points need to be raised and answered.

- *What research themes got most and least attention in the last 3 years?* By interpreting Figure 5, based on a combination of publications years and sub-themes since 2012, most of the studies focused on enabling *distributed resource access* [S20, S26, S27, S56] for supporting robotic coordination and architectural *design* of robotic development [S33, S40, S51, S55]. There is no evidence for the research on *reconfiguration, reengineering* and *programming* for robotics.

- *What years are reflected as most progressive based on the number of research publications and what was the focus of research during those years?* By analyzing temporal distribution facet (x-axis left), we can identify 2006 to 2014 as being most progressive years in terms of total publications (a total of 42 studies, 75% approx. of the total). During these years, the studies focused all identified themes and their sub-themes except for the research on architecture-based reengineering of robotics software. The major focus of research during these years was on architecture-driven *reconfiguration* [S32, S36], *design* [S17, S43] and *modeling* [S11, S33] of robotics software.

# 5. Architectural Solutions and Frameworks for Robotic Software

Based on a taxonomical classification of existing research, we specifically focus on answering **RQ 1.2** that investigates: *What types of architectural solutions have been reported for robotic software?* (Section 5.1) *and* **RQ 1.3**: *What architectural frameworks have been provided to support the solutions?* (Section 5.2).

## 5.1 Problems and Architectural Solutions for Robotic Software – An Overview

Table 2 serves as a catalogue that provides references to problem-solution mapping by highlighting the (i) recurring *problems*, while (ii) *solution* presents architecture-centric approaches to address the problems. The presentation of related studies complements the view by highlighting the relevant studies as an evidence of problem-solution mapping. For each of the reviewed studies, the problem and solution views were captured in data extraction form - **Appendix B** (M01, M02 in **Table IV**) and summarized in Table 2. Whilst the thematic classification (borrowed from Figure 4) organizes similar or distinct problems, whereas sub-themes represent specific type of problems. For example, in the context of *Robotic Adaptation and Reengineering* Table 2 highlights that:

- **Problem view**: *how to reconfigure a robot's components (and behavior) such that it can adapt itself to operate optimally with available resources and evolving requirements of its environment?*

- **Solution view**: *the proposed architectural approaches to address this problem are model-driven and policy-based adaptation of robotic software.*

- **Related studies**: *the related studies supporting the above mentioned solutions are* [S5, S7, S23, S30, S32, S36, S39, S41, S42].

Table 2. Problem Solution Mapping with regards to Generic and Thematic Classification of Research.

| Operations of Robotic Software | | |
|---|---|---|
| *Thematic Classification* | *Problem –Solution View* | |
| | **Problem** | **Solution** |
| *Robotic Coordination* | *Type*: **Information Fusion**<br><br>*Challenge*: How to support the collection, processing and sharing of information that is fused into a team of robots to support their operations? | *Approach*: **Client-Server Model**<br><br>*Solution*: Intra-robot interaction is enabled by collecting contextual and mission-specific information from individual robots (*mission agents*) and sharing it through a central node (*mission server*). |
| | Related Studies [S1, S9, S35, S47, S54] | |
| | *Type*: **Distributed Resources Access**<br><br>*Challenge:* How to enable the access and utilization of distributed resources (hardware components, that are often virtualized) to assemble/operate a robot? | *Approach*: **Service Oriented Architecture**<br><br>*Solution*: Exploit distributed and loosely coupled services that can be dynamically discovered and invoked. A collection of software services can be mapped to one or more hardware resources. |
| | Related Studies [S6, S15, S20, S26, S27, S56] | |

| | *Evolution of Robotic Software* | |
|---|---|---|
| | **Problem** | **Solution** |
| *Robotic Adaptation and Reengineering* | *Type*: **Reconfiguration** <br><br> *Challenge:* How to reconfigure a robot's components and behavior such that it can adapt itself to operate optimally with available resources and evolving requirements? | *Approach*: **Model-driven and Policy-based Adaptation** <br><br> Solutions <br> - *Model-driven adaptation*, architecture model is transformed dynamically to support transformation-driven adaptation (models@runtime) <br><br> - *Policy-based approaches* directly or indirectly rely on the IBM framework for Monitoring, Planning, Analyzing and Executing (MAPE loop) to support adaptation tasks. |
| | Related Studies [S5, S7, S23, S30, S32, S36, S39, S41, S42] | |
| | *Type*: **Fault Tolerance** <br><br> **Challenge**: How to enable a robot to continue with its operations despite the presence of failure? | *Approach*: **Fault Tolerant Middleware** <br><br> *Solutions*: A middleware layer is provided that supports a minimum of three fundamental operations as i) context monitoring, ii) fault identification, along with fault ii) recovery and minimization. |
| | Related Studies [S4, S8, S48] | |
| | *Type*: **Reengineering** <br><br> *Challenge*: How to support re-engineering of an existing robot to an evolved robot that better satisfy new requirements? | *Approach*: **Architectural Refactoring** <br><br> *Solutions*: A conventional 3-step process that requires i) recovering, ii) refactoring, and iii) optimizing the legacy architecture towards an evolved architecture |
| | Related Studies [S14] | |
| | *Development of Robotic Software* | |
| | **Problem** | **Solution** |
| *Robotic Modeling, Design and Development* | *Type*: **Modelling** <br><br> *Challenge*: How to exploit high-level models that take a step from code-based to model-based development of robotic systems? | Approach: **Model-driven Development** <br><br> *Solution*: Exploits the high-level models to capture the domain requirements that are used to generate the code. Architectural components bridge the gap between the requirements and executable source code. |
| | Related Studies [S11, S13, S38, S40, S44, S51, S55] | |
| | *Type*: **Design** <br><br> *Challenge*: How to abstract the modules of code and their interconnections as architectural components and connectors to support the notion of component-based robotics? | *Approach*: **Component-based Robotics** <br><br> *Solution*: Architectural components are developed or reused to abstract the source-code level complexities for component-based development. These are also referred to as *COTS (Component Off-The Shelf) based robotics*. |
| | Related Studies [S2, S16, S17, S18, S22, S24, S25, S28, S29, S31, S34, S37, S43, S45, S49, S50, S52, S53] | |
| | *Type*: **Programming** <br><br> *Challenge*: How to develop and extend the frameworks that enable reuse and modularity for robotic programming? | *Approach*: **Object and Module-oriented** <br><br> *Solution*: The solution provides frameworks to abstract design notation (typically object-oriented modeling) to support reusable and modular programming. |
| | Related Studies [S3, S10, S12, S13, S19, S21, S46] | |

## 5.2 Framework Support for Architecture-driven Solutions

We report the identified architectural frameworks and their support for robotic software to answer RQ 1.3. An architectural[5] framework facilitates a diverse set of architecting activities (e.g., modeling to development and documentation) to promote architecture-centric solutions. Based on the analysis of architectural solution (RQ 1.2), we present the architectural frameworks for robotic systems that appeared in at-least two studies. In Table 3, we do not include any solution specific or once-off frameworks (such as SAW - Surgical Assistance Workbench for developing surgical robots [S23]). We present the key aspect of four architectural frameworks in Table 3, while the discussion is guided based on frameworks overview in Figure 6.

---

[5] SEBoK Architecture Framework (glossary) "An architecture framework establishes a common practice for creating, interpreting, analyzing and using architecture descriptions within a particular domain of application or stakeholder community. Examples of architecture frameworks: MODAF, TOGAF, Kruchten's 4+1 View Model, RM-ODP. (ISO/IEC/IEEE 2007)"

Table 3. A Summary of identified Architectural Framework for Robotic Software

| Frameworks | | | | |
|---|---|---|---|---|
| *Intent* | *Support Activities* | *Source Type* | *Support for Systems* | *Studies* |
| | | | | |
| **Component for Robotics (Orca)** | | | | |
| - Component Distribution<br>- Component Reusability<br>- System Modularity | Development | Open | Robotics | [S1, S32] |
| **Open Platform for Robotic Service (OPRoS)** | | | | |
| - Component Composition<br>- Component Execution<br>- Component Life-cycle Management | Development | Open | Robotics | [S4, S31] |
| **Programming in Small and Many (PRISM)** | | | | |
| - Distributed Component Execution<br>- Resource Efficiency<br>- Dynamic Reconfigurability | Evolution | Closed | - Hand-held Devices<br>- Embedded Systems | [S5, S30, S39] |
| **Healing, Adaptive, and Growing SoftwarE (*SHAGE*)** | | | | |
| - Self-healing<br>- Acquisition and Application of Adaptation Knowledge | Evolution | Closed | - Robotics<br>- Self-adaptive Software | [S23, S24, S36] |

To answer RQ 1.3, we only highlight the core features of the frameworks and their impact on robotic systems. The extended details about individual frameworks can be found in relevant studies (highlighted in Table 3). The frameworks in Table 3 act as middleware or an abstraction layer to hide the platform and implementation specific complexities to provide an environment for the development and execution of architectural components for robots. Table 3 presents (i) primary intent of the frameworks, (ii) the types of activities a framework supports (based on *Generic Classification* of identified research themes, Figure 4), (iii) source type of the framework as either open or closed source project, (iv) type of systems they support and v) the related studies that have utilized these frameworks. In recent years, an increasing distribution and utilization of open source software or specifically frameworks for robotic systems [S1, S4, S31, S32] is progressing the research and practices to a model in which code is distributed, repeatedly executed and built upon. The open-source robotics software accelerates robotics development as community-wide efforts by developing, sharing and optimizing robust algorithms, algorithm comparison, and collaborations between research groups and robotic software developers [63].

Figure 6 is a collection of diagrams adopted from their respective sources ([S5, S23, S31, S32]) and has two distinct purposes while answering RQ 1.3. Specifically, Figure 6 provides (i) a visual representation of each framework in terms of individual elements and their relations that constitute a framework, and (ii) illustrative examples of the operations and application of the framework. For example, in Figure 6 A) an overview of OPRoS framework highlights that framework provides an environment (acting as a container) for composition and execution of various architectural components (A, B, C). Figure 6 and Table 3 are complementary, Table 3 highlights the key attributes of and Figure 6 explains the frameworks.

1. **Orca** as an open-source framework implements a component model to support component-based robotics. As illustrated in Figure 6 a), the components of Orca framework are deployed across two different robots (Robot X and Robot Y). The modularity of system is maintained by encapsulating specific functionality in an appropriate component that can be utilized and reused through components coordination. Figure 6 a) illustrates that by exploiting components, Orca implements i) interfaces (component's provided and required functionality), and ii) infrastructure to enable interface binding (component messaging). The benefits of Orca components distribution are demonstrated in [S1] to support coordination between a team of robots (i.e., Figure 4 – Taxonomical Classification).

2. **OPRoS** is also an open-source and component-based framework to develop robotic systems. The framework consists of multiple components and acts as a process in an operating system (e.g., Microsoft Windows and Linux) for component-based development and operations of a robot as in Figure 6 b). First, the composer composes a number of components (having ports and connectors) that allows the execution engine to start executing the components to support robotic operations such as sensing and navigation. As illustrated in Figure 6 b), the framework acts as a container and provides the execution, lifecycle management, configuration, and communication of various components. In the review, the studies [S4] and [S31] have utilized OPRoS that supports evolution (with fault tolerant robotic [S4]) as well as the development (with component-based robotic [S31]). It is vital to mention that OPRoS also provides an environment for robotic simulation by utilizing (i) a physical engine that simulates the robot and (ii) a graphics engine to simulate the objects and obstacles for the robot. The simulation feature offered by OPRoS helps robotic software developers to not only develop but also evaluate the solution by means of simulations before deployment [S31].

3. **SHAGE** as a framework consist of two core elements (i) *modules* (inner or white area in Figure 6 c)) as a collection of components supporting various functionalities and (ii) *repositories* (outer or blue area in Figure 6 c)) for data acquisition and utilization. The framework supports a dynamic adaptation of architecture. The modules support various tasks such as component brokerage, reconfiguration and decision making to address the issues of *what to adapt*? In Figure 6 c), the two circles represent external elements as *Environment* and *Human* (that are outside the framework) but has an impact on the adaptation process. For example, the environment (e.g., navigation path) on which a robot operates imposes certain adaptation constraints (i.e., obstacle avoidance) that must be accommodated by the framework. Therefore, SHAGE also enables a communication with the robotic environment and user elements to decide on *what and when to adapt*? In the review, the studies [S23, S24, S36] exploit the SHAGE framework. Two studies [S23, S36] are specifically focused on reconfiguration; one study [S36] supports the development of an adaptive robot.

4. **PRISM** or sometimes also referred as Prism-MW (Prism-Middleware - implemented in Java and C++) is a framework providing middleware support for an efficient implementation, deployment, and execution of architectural elements. A layered overview of PRISM is presented in Figure 6 d) that abstracts the hardware. We discuss the architectural middleware and advanced services (top two) layers that are more relevant to robotic software than the underlying layers. Figure 6 d) presents the view where architectural middleware layer abstracts the operating system and virtual machine details to provide an environment for the utilization of architectural elements including components, connectors, events, and ports. Advanced services layers support the deployment, monitoring, adaptation and other functionalities for architectural elements. Unlike Orca and OPRoS that are robot specific platform, PRISM enables the development of hand-held and embedded systems thus making robotics as one specific domain that can utilize this framework. We identified three studies [S5, S30, S39] that utilize PRISM to support runtime evolution by means of dynamic reconfiguration of a robot. Two studies [S30, S39] utilize PLASMA [S39] framework that offers a plan-based adaptation of architectures but PLASMA itself extends PRISM to support adaptation plans on architectural component and connectors.

It is important to discuss a couple of frameworks that are not included in Table 3 but were found in the reviewed studies. These frameworks are CLARAty and MRDS and ROS. CLARAty, Coupled Layer Architecture for Robotic Autonomy (CLARAty), [S2] has been developed at Jet Propulsion Lab for improving the modularity of software by NASA to support robotic autonomy for space exploration. Microsoft's Robotics Developer Studio (MRDS) is a platform-dependent environment for robotic control and simulation. One study [S9] utilizes MRDS's runtime support for services orchestration for developing service-driven robots.

The Robot Operating System (ROS) [S27] is not an operating system in a traditional sense, however, as open source platform/framework ROS provides a structured communications layer/middleware above the operating systems to support the development and operations of robotic systems. In contrast to CLARAty [S2] and MRDS [S9], ROS represents the recent - community-driven efforts – for open-source and collaborative development of robotic software [62, 63].

**a) Component Distribution and Coordination in ORCA Framework**

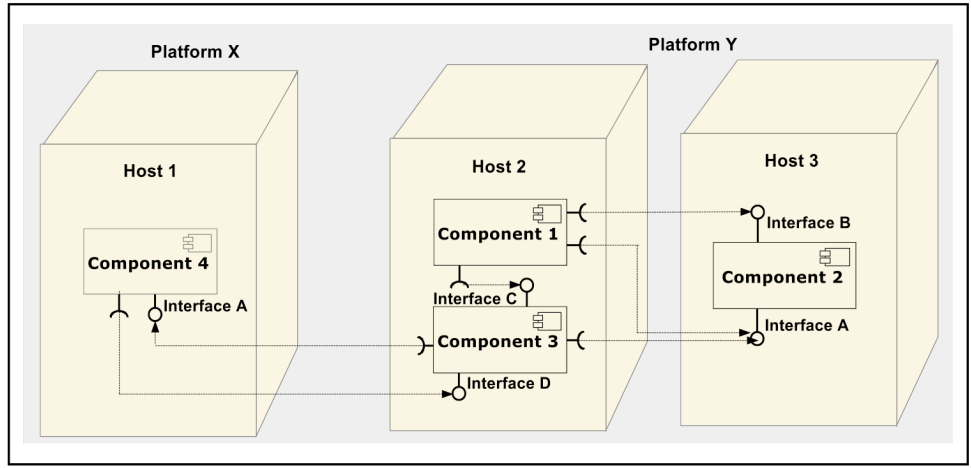

**b) Component Composition and Execution in OPRoS Framework**

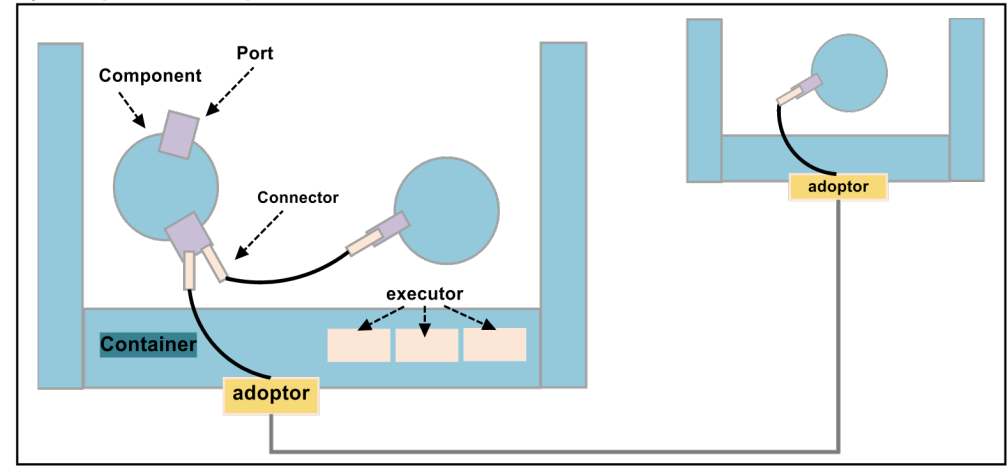

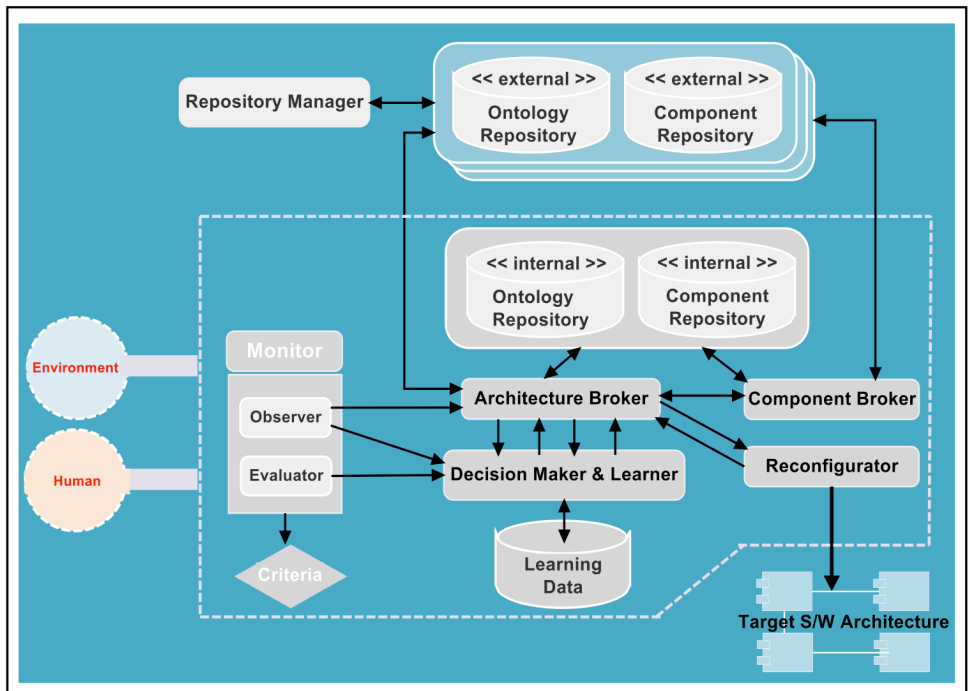

**c) Layered View of Repositories and Modules in SHAGE Framework**

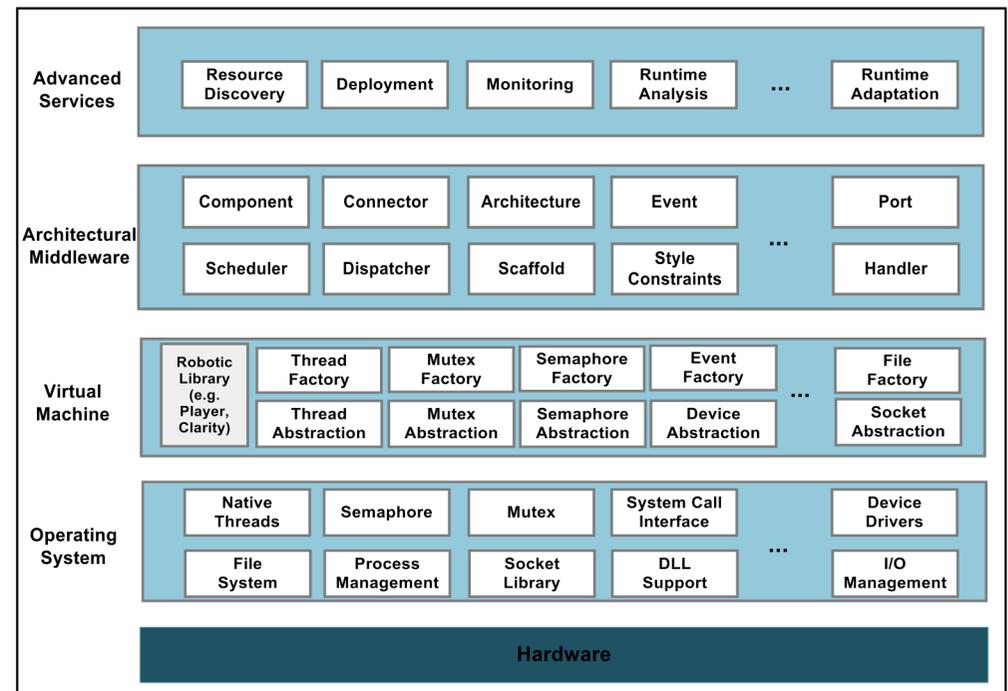

**d) Layered Representation of Components in PRISM Framework**

Figure 6. An Overview of the Various Architectural Frameworks for Robotic Systems.

# 6. Notations, Validations and Domains of Architectural Solutions

We now discuss the aspects of architectural solutions for robotics systems (i.e., architectural notations, validation methods and application domains of the solutions (cf. section 5)) to answer RQ 1.4 – RQ 1.6. We aim to investigate: *RQ 1.4: What architectural notations have been exploited for solution representation?* (Section 6.1) *RQ 1.5: What validation methods have been employed to evaluate the solutions?* (Section 6.2) *RQ 1.6: What were the application domains for architectural solutions?* (Section 6.3). The answers to these questions conclude the findings for RQ 1 (state-of-research) and also enable us to answer other RQs later.

## 6.1 Architectural Notations for Robotic Systems

One of the primary benefits of architecture-centric development of a robotic system is related to abstracting the complex implementation specific details with architecture-driven modeling of software [18, 22]. Architectural modeling provides a global view of software (robotic components and their connectors) [21] by abstracting lower-level details (such as manipulator or controller level code) [12]. We briefly discuss the reported architectural notations utilized for robotic software to answer RQ 1.4. The term notation refers to models/descriptions for architectural representation such as Unified Modelling Language (UML) [S23], Architecture Description Language (ADL) [S41] or similar models such as ontology based representation [S24, S55] as illustrated on Figure 7. For example, Figure 7 A) [S41] indicates the application of an ADL for architecture representation. Figure 7 also provides two-level details: (A) an overview of all the identified architectural notations and (B) UML-specific notations as they represent an overwhelming majority with 63% of the entire notation (identified in 35 studies).

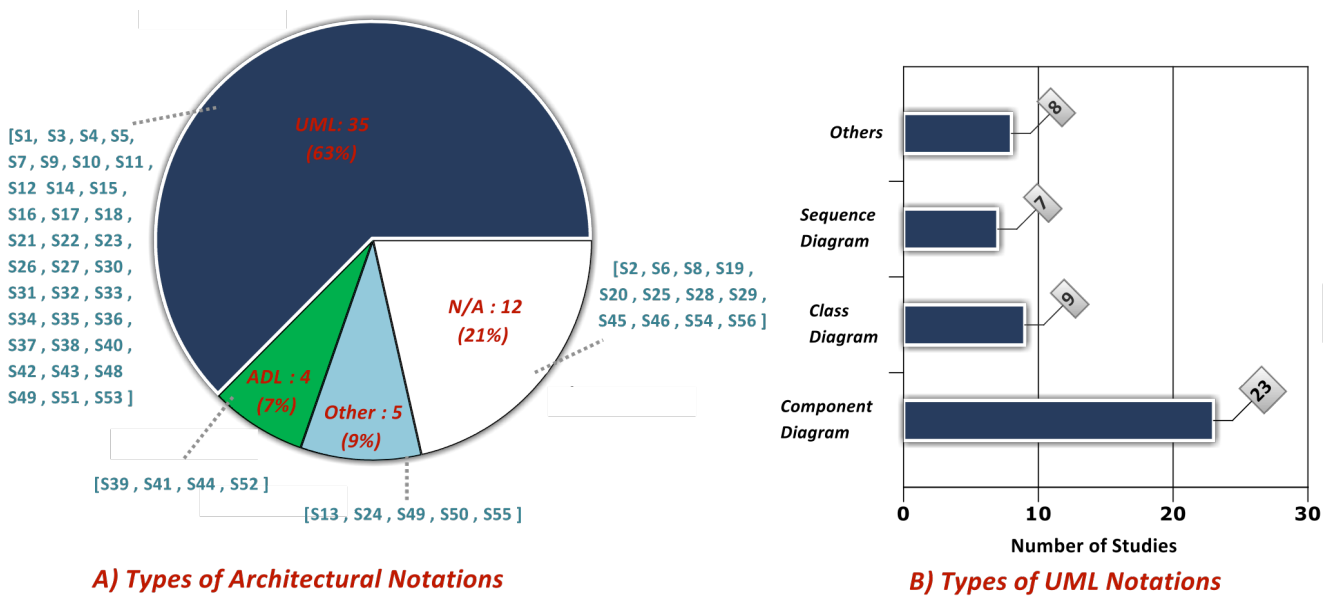

*A) Types of Architectural Notations*

*B) Types of UML Notations*

Figure 7. Overview of Architectural Notations for Modelling Robotic Software.

*UML-based Notations* have been used in the majority of the research solutions reviewed in this study. UML is an intuitive and widely adopted graphical notation for architecture modelling that might have been found easy to understand and use by robotic researchers. Another advantage of UML being industry standard for robotic development [47]. In this context, the study [S24] demonstrates the feasibility of UML (with structural and behavioral modeling) of a home-service robot named T-Rot. We provide an overview of the specific UML notation (identified in at-least 3 or more studies) in Figure 7 B). Figure 7 A) also shows that some of the studies like [S34, S37] utilized multiple UML notations (e.g., state-chart and sequence diagram) and each notation from respective study is counted once to derive Figure 7 B). The component (23 studies) and class diagrams (9 studies) are prominent choices and fundamental notations for modelling robotic components - represent an overall structure of a robot system. In comparison, for behavioral modelling - component interactions such message passing between actuators and sensors [S35, S36] – sequence diagrams are utilized (7 studies). Moreover, the categorization named "others" represents occasional notations such as activity [34], state [S7] and deployment diagrams [S9] used in specific and solution dependent scenarios for robotics software.

***ADL-based Notations*** have been used in 4 (7%) of the reviewed studies - exploiting ADLs for dynamic adaptation [S39, S41] and aspect-oriented development of robotics [S44]. One study [S39] specifically extends C2SADEL with event-based architecture style for plan-based adaptation of robotics using PLASMA framework. Another study [S41] extends the xADL 2.0 schema for a policy-based adaptation of robotics. An interesting discussion about the practical implications and limitations of plan versus policy-based adaptation of robotics is provided in [S39]. The intent of both studies [S39, S41] is similar i.e., to enable architecture-level adaptation (runtime evolution) of robotics. However, the solution specific needs (plan versus policy based adaptation) determine the needs for selection of an appropriate ADL. The review suggests the preference for UML models compared with ADLs is determined by its ease of use both in the context of academic research and industrial solutions [47]. A recent industrial survey on architectural languages [48] also highlights that academic ADLs seem not to fulfill the industrial requirements. The survey also suggest that academic research and development of ADLs can provide some inspirations to develop and enhance industry specific ADLs.

***Other Notations*** are used in 5 (9%) studies for a variety of purposes and solution-specific needs. For example, [S13] utilizes the module diagram to represent an object-oriented architecture for a surgical robot, named ROBODOC. The preference for modular representation is determined by a mapping of (design elements) modules and their interactions to (source code elements) objects and their relations. UML was not available at the time as the study was published in 1991. Park et al. [S24] reported an ontology-based modeling of architecture for the actions of a robot. The ontology-based modeling facilitated the structural and semantic matching and searching of appropriate actions on architecture models of robotic software.

***No Explicit Notations*** have been presented in 12 (21%) studies as illustrated in Figure 7 A). For example, one study [S6] provides a generic representation of SOA style (based on service provider, publisher and requester); however, no explicit details about modeling individual services have been provided. Another study [S19] utilizes the design patterns (black-board and message-broker) for architectural design of mobile robots, however, the study uses only generic boxes and arrows for architectural descriptions, rather than a specific notation such as the component diagrams.

These findings indicate that robotic research on architectural solutions heavily relies on UML models. The 4 studies that used ADLs have been reported by software architecture researchers, who are expected to have more knowledge about ADLs. Non-software engineering researchers are more inclined to use UML models. In this context, a study [48] suggests that ADLs should have features to support communication among different types of stakeholders about individual activities of architecting process. UML models despite their simplicity offer diverse modeling notations (structural, behavioral, and interaction diagrams) to address (i) different activities such as structural development, behavioral adaptation and (ii) accommodate more stakeholders such as robotics/artificial intelligence/industrial engineering and automation researchers, developers and other practitioners.

## 6.2 Solution Validation and Architectural Evaluation Techniques

We discuss how the architectural solutions have been evaluated to highlight the relative strength and validity of the evidence to answer RQ 1.5. We distinguish between solution validation and architecture evaluation as validating the overall solution aimed to analyze the ability of a robotic (software) system to address the identified problems [20, 37]. In contrast, *architectural evaluation specifically focuses on the ability of an architecture (as a part of overall solution) to satisfy - functional and non-functional - requirements* [38]. In Figure 8 A), first we highlight how the validation results have been presented that follows a discussion of the solution validation and architectural evaluations in Figure 8 B). Figure 8 A) illustrates three distinct types as:

(i) validation of the overall solutions (59%), (ii) architecture-specific evaluation (14%), and (iii) implementations and lessons learned (27%). Implementations and lessons learnt refers to implementing a preliminary solution (prototype or proof-of-concepts) and report the lessons learned to guide future developments. These have been represented in a total of 15 studies as illustrated in Figure 8 A). For example, a study [S25] presents the lessons learnt in the development and preliminary evaluation of Service Oriented Robotic Architecture (SORA) for space exploration mission. The study highlights the benefits (e.g., component distribution and reusability) and their associated challenges (middleware requirement) of exploiting the concepts of service-orientation or service-oriented robotics [18, 14] for mission critical tasks. Other lessons include the feasibility of programming models and components to support development, or evolution (reengineering [S14] and adaptation [S24]) of robotic systems.

We have identified four distinct validation methods that explicitly discuss architecture-specific evaluations. Figure 8 A) provides a percentage comparison of *solution validations* versus *architectural* evaluations. Figure 8 B) focuses on solution validation versus architectural evaluations in the reviewed studies.

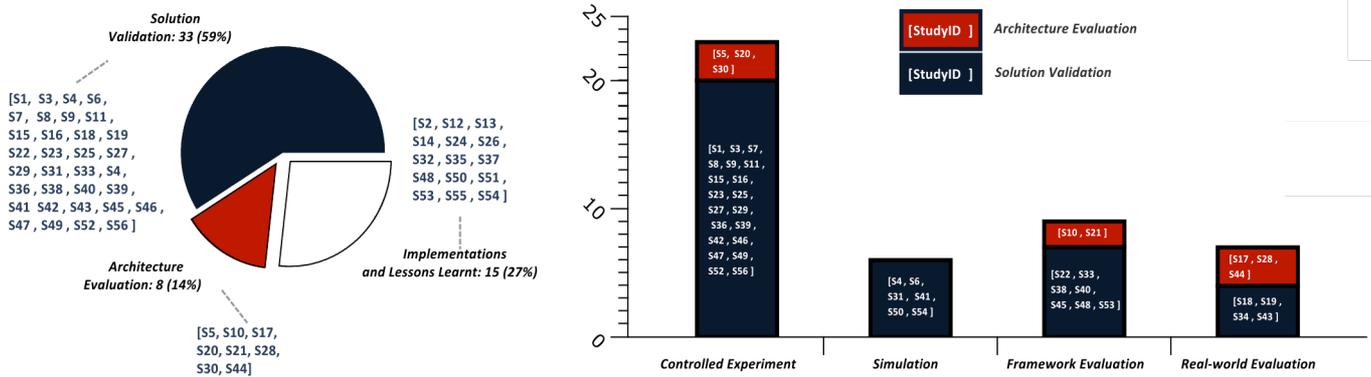

*A) Presentation of Results*         *B) Solution Validations and Architecture Evaluation*

Figure 8. Overview of Solution Validation and Architectural Evaluations.

***Controlled Experimentation** methods are focused on laboratory-testing to validate or experiment with the (partial) solutions in controlled environments.* Controlled experiments have been used in a majority of the studies (23 in total, 41% approx.). For example, one study [S39] has used controlled experiment to validate the adaptation of navigational robots (e.g., single robot [S1] and team of robots [S5]) on a predefined or fixed path. Early experimentation is fundamental to establish benchmarks and to seek preliminary feedback for solution refinements or further commercialization of robotic solutions [35, 39]. **Architecture-specific evaluations** have been reported in a limited number of studies.

One study [S5] evaluates the overall solution to support robotic navigation and then also compares and evaluates the proposed PRISM-MW with two reference architectures (i) PBAAM [S26] and (ii) layered adaptive architecture [40]. The studies [S20, S30] use empirical evaluation of the architectural solutions. Specifically, the study [S20] evaluates a novel cloud-based architecture by analyzing the energy consumption and computational off-loading of robotics architecture – highlighting the feasibility of cloud robotics in resource constrained environment. In comparison, [S30] utilizes a conventional metric called Constructive Cost Model (COCOMO) for analyzing architectural support to develop solution with reduced efforts from months to weeks (approximate reduction from 4 – 9 months to 4 weeks).

***Simulation** refers to establishing and exploiting a virtual environment (as closely mapped to real world context, as possible) to evaluate the functionality of the solution.* Simulations do not reflect all the possible scenarios that can be encountered in a real world, however, it projects many of the same characteristics as multiple sources of input to decide which actions to perform for robots (6 studies, 11% approx. of studies in Figure 8 B)). The simulation-based validations are conducted with two of the well-known frameworks OPRoS (Open Platform for Robotic Services also highlighted in Section 5, cf. RQ 1.3) and ROBOCODE that simulate robotics navigation. The OPRoS consists of a physical engine to simulate the robot and a graphics engine to simulate the objects and obstacles in the solution [S4]. In comparison, a specific example of ROBOCODE is the simulation of the battle-field to analyze the adaptation of robotics [S41]. We did not find any evidence of simulation-based methods to specifically evaluate the architecture of robotics software.

***Framework Evaluation*** *aims at evaluating the capability or effectiveness of newly developed or already existing frameworks that support development of robotic systems.* These frameworks are mostly evaluated to support the reusability and modularity (e.g., through programming [S10], componentization [S22] or modeling [S33]) to support architecture-based development of robotic systems. These represent a total of 9 (16%) studies. ***Architecture-specific Evaluations*** are supported such as evaluating Robot Independent Programming Language (RIPL) [S10] to support development and reusability of components for programming robotic systems. The study reported in [S21] evaluates the application of patterns to support component-based robotic systems [13, 21].

***Real-world Validations*** evaluate the solutions in a real world with realistic scenarios and environments. The validations are in the context of commercial robots that support industrial automation [S17, S28], surgical procedures [S18], and autonomous vehicle [S19, S43, S44]. Architecture-specific evaluations are only reflected in a limited number of studies focused on the degree of reusability of various architectural components [S28, S44] and evaluating the real-time execution of the components [S17].

A majority (33, 59%) of the studies focus on validating the overall software solution, while neglecting architectural evaluation. A lack of architecture-specific evaluations (14% studies) can be interpreted as a weakness in the evidence for the effectiveness of architectural solutions for robotic systems. In the context of Figure 8 B), a lack of architecture specific evaluations hinders the validation of non-functional or quality requirements [38]. The work in [60] can be viewed as an inspiration for the future research that aims to evaluate the architectures of mobile robots against the attributes such as portability, ease of use, software characteristics, programming and run-time efficiency to qualitatively validate a robotic system.

## 6.3 Application Domain

We also identified the domains to which the architectural solutions had been applied [7, 10] to answer RQ 1.6. The domain or application context of a robotic system defines*: the environment (real world or a simulations) in which a system operates under some specified conditions (to satisfy some requirements)* [21, 24]. A typical example of such a domain is mission control in which robot(s) must support mission critical operations such as space exploration [S2, S26] or resolving some emergency scenarios [S9]. Figure 9 presents a relative distribution of the types of domains to which architectural solutions have been applied.

***Mission Control Robotics*** *represent a type of mission critical systems [30] that are developed to accomplish a specific mission identified in a total of 10 studies (18%)*. For example, the study [S9] represents (first responder emergency scenario) as type of a safety critical system, while [S13] (surgical robot) operate in a domain with health and safety criticality. The reviewed studies [S2, S26] suggest as early initiatives by NASA to develop robotic systems for space exploration. The space exploration robots represent earlier (prototype-oriented) solutions that require an appropriate human intervention or supervision to carry out their mission. We interpret Figure 9 as: mission critical robots are discussed in 10 (19% of studies), and mission/system criticality [31] are classified as: (i) *space exploration* [S2, S26], (ii) *hazardous waste cleanup* [S8, S10, S43, S44, S47], (iii) *first responders* [S9, S11], and (iv) *simulated combat* [S41] robotics. Early experimentation and lessons learned [S26] provide the guidelines and architectural needs to develop futuristic robots that assist NASA with their space exploration program[6] [32].

***Service Robotics*** *are the type of robots that replace humans' services (such as cleaning [S42] and home service [S14]) for common domestic tasks.* These robotic systems represent a relative majority of the reviewed literature (13 studies, 24%). Recently, several domestic service robots are increasingly considered as consumer products with the primary aim to support and increase the quality of life in many areas [33].

This has also prompted a number of renowned consumer product companies (Sony [2], Honda [1], and Samsung [S14]) to invest in research and development of service robots. The home service robots range from experimental solutions (academic research) with path navigation and object pickup capabilities [S3, S6] to commercial solutions offering drink serving [S7, S42], clean-up [S42] and home surveillance solutions by Samsung robotics [S14].

---

[6] Robotics|NASA: http://www.nasa.gov/education/robotics/#.U8O_9_mSzyl

***Navigation Robotics*** *also known as mobile robots that represent a class of robots that move autonomously from its current (source location) position towards a goal position (target location), while avoiding situations such as collisions and obstacles.* Most of the identified studies have focused on experimental navigation with a robot tracking certain objects or patterns of unknown paths [S25, S38]. A concrete implementation of robotic navigation is in home service robots where navigation is fundamental requirement to provide the services such as home clean-up [S42]. Another experimental case is leader-follower navigation by robots [S5, S30, S39] where instead of a single robot navigation, the follower must keep track of their leader and navigate accordingly to collect and share the environmental and contextual information. A specific example of navigational robotic is the autonomous vehicle [S19, S50] that can drive itself on a given path by following traffic lights and obstacle avoidance. This concept has gained a lot of attention and finding its application in various dimensions, specifically for the construction of military and commercial automobiles [35, S50, S54].

***Industrial Automation Robotics*** *or industrial robotics (based on ISO 8373 standardization) aim to automate complex industrial tasks such as assembling, painting, testing and packaging of industrial products with speed and precision [34].* The studies [S21, S29] represent experimental solutions. A study in [S10] represents a real-world robotic application for holding/un-holding and packaging the bulk of industrial material. The studies [S28, S23] automate the assembling of the industrial products.

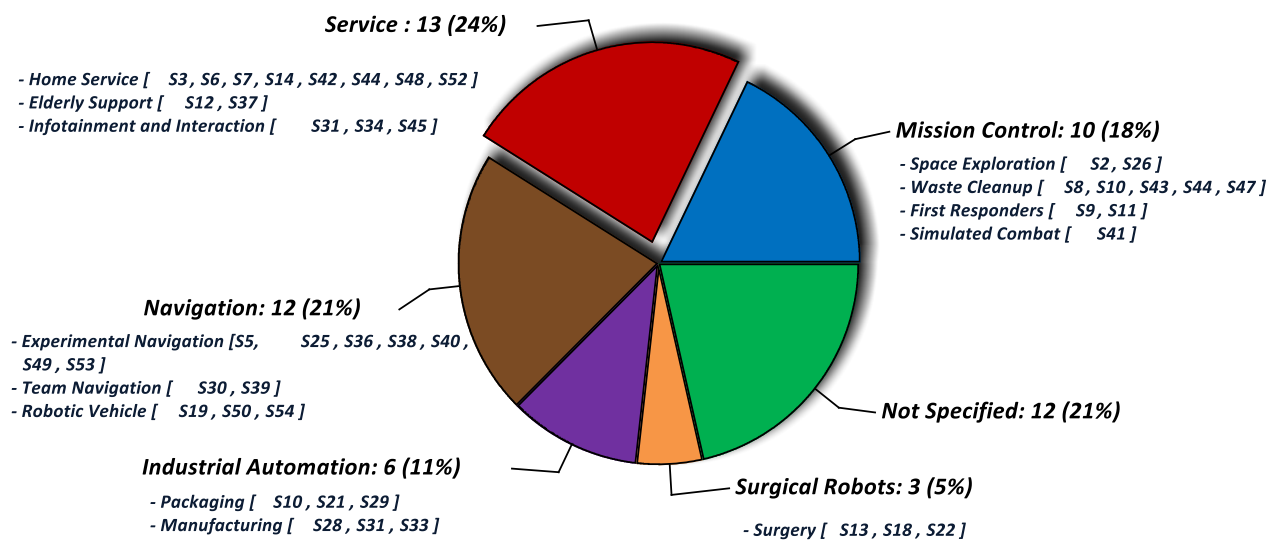

**Service : 13 (24%)**
- Home Service [  *S3, S6, S7, S14, S42, S44, S48, S52* ]
- Elderly Support [  *S12, S37* ]
- Infotainment and Interaction [  *S31, S34, S45* ]

**Mission Control: 10 (18%)**
- Space Exploration [  *S2, S26* ]
- Waste Cleanup [  *S8, S10, S43, S44, S47* ]
- First Responders [  *S9, S11* ]
- Simulated Combat [  *S41* ]

**Navigation: 12 (21%)**
- Experimental Navigation [S5,  *S25, S36, S38, S40, S49, S53* ]
- Team Navigation [  *S30, S39* ]
- Robotic Vehicle [  *S19, S50, S54* ]

**Industrial Automation: 6 (11%)**
- Packaging [  *S10, S21, S29* ]
- Manufacturing [  *S28, S31, S33* ]

**Surgical Robots: 3 (5%)**
- Surgery [  *S13, S18, S22* ]

**Not Specified: 12 (21%)**

Figure 9. Study Distribution by Application Domain of Robotics.

***Surgical or Medical Robots*** *support the notion of minimally invasive surgery (MIS) where a robot plays the role of a surgeon or assists a surgeon to carry out the surgical procedures [36].* We found three studies with architectural support for surgical robots. Two studies [S13, S18] represent a robotic system to perform the planning, execution of surgery with manual takeover if required. One solution [S22] allows integration of various components (sensors and imaging devices) that help a surgical system to function.

Our findings enable us to conclude that the growing complexity of modern and practical robots increase the significance and importance of good software architecture design for reliably and efficiently developing and evolving commercial robotic systems [35]. The state-of-research represents early solutions or prototypes that may have a significant impact in the near future on robotic driven world such as robotic surgeons [S13] or unmanned [S54] and self-driving vehicles [S50]. Moreover, the mission critical robotics are emerging as human alternate to execute the trivial and complex tasks such as hazardous waste cleanup [S8], rescue mission [S9] and space exploration [S2]

# 7. Demography of Published Research

This Section presents and discusses the demographic details of the reviewed studies in order to answer RQ 2 (i.e., the publication years, frequency, sources and active communities of the research). We aim to address the following three sub-questions: *RQ 2.1: What is the publication frequency and fora of research over the years?* (Section 7.1), *RQ 2.2: What are the prominent venues of publication and the types of research published?* (Section 7.2), and *RQ 2.3: What research communities are active on software architectural solutions for robotic systems?* (Section 7.3). The discussion of results in the remainder of this section also guides our discussion about research progression and future dimensions later in this paper.

## 7.1 Years and Types of Publications

Figure 10 shows the years and types of publications of the reviewed studies to answer RQ 2.1. Figure 10 A) shows the relation between the total numbers of studies (y-axis) published during individual years (x-axis) since 1991 to 2015 (until August). Figure 10 represents a bar graph, where each bar shows a relative distribution of the different types of publications (*Book Chapters, Journal, Conference and Workshop and Symposium Papers*). For example, the bar relative to year 2007 represents a total of 06 studies published (publication distribution as: *Conference Papers: 03*, *Journal Paper: 01* and *Book Chapters: 02*) in that year. The initial studies (such as [S10] and [S13]) were focused on exploiting object-oriented design and framework to enable modular and reusable programming of robotic systems. Figure 10 indicates that, no study from 1993 to 1997 has been included in our review because: (i) our literature search did not produced any relevant results, or (ii) the studies did not passed the qualitative evaluation from those years.

From 1998 to 2003, only three studies [S8, S2, S28] were published that moved the research beyond the robotics programming to focus more on component-driven development. Since 2004 to 2015, there has been a noticeable increase in the number of studies on architectural solutions for robotic systems. There were 50 studies (90%) published from 2004 to 2015 that represents the research progression in the last decade.

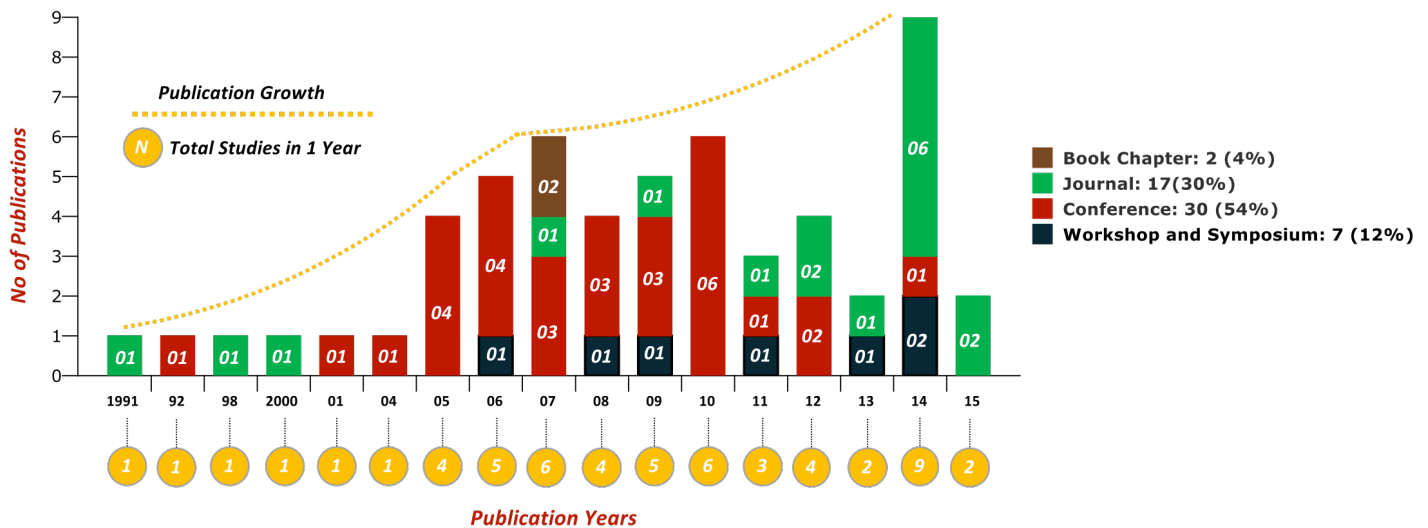

Figure 10. Overview of the Publication Years and Types

The solutions reported in these studies mainly focus on component-based [21] and service-driven [18] robotics. The research on service-driven robotic systems [S20, S27] has been mainly published from 2012 to 2014. The focus of the studies published in recent years is on *fault-tolerant* [S48], *cloud-based* [S55, S56] and *model-driven robotics* [S51]. The number of conference papers represents approximately 54% of all the publication types; the percentage of journal, symposium and workshop papers is 30% and 12% respectively.

We also reviewed two book chapters [S45, S46] published in 2007. The book itself [21] provides interesting insights about addressing various software engineering challenges for robotic systems.

## 7.2 Main Sources by Publication Frequency

We report the main sources (publication venue/sources) of the reviewed studies based on publication frequency as presented in Table 4 to answer RQ 2.2. such information is expected to help us to identify the representative research communities and an evidence of robotics or specifically software architecture for robotics as a multi-disciplinary research. We present the prominent publications venues in terms of journals, conference series or symposium/workshop along with the type of research that has been published in those venues. We only highlight the publication sources that have at-least two or more studies published as in Table 4. We do not provide extensive details of the published studies (already discussed in Section 5, Table 2). Table 4 indicates the studies (*Study ID*) that provide a reference to consult previous sections. Table 4 highlights a mix of robotics and software engineering research venues and their research focus.

Table 4. Overview of Main Sources by Publication Frequency

| Study ID | Publication Source | Type | Acronym | Community |
|---|---|---|---|---|
| [S47, S48, S49, S56] | Robotics and Autonomous Systems | Journal | RAS | Robotics |
| [S22, S25, S32] | IEEE/RSJ International Conference on Intelligent Robots and Systems | Conference | IROS | Robotics |
| [S5, S36, S41] | International Workshop on  Software Engineering for Adaptive and Self-Managing systems | Workshop | *SEAMS | Software Engineering |
| [S8, S28, S31] | IEEE Transactions on Automation Science and Engineering | Journal | **T-ASE | Robotics |
| [S14, S37] | International Conference on Software Engineering | Conference | ICSE | Software Engineering |
| [S50, S54] | Journal of Software Engineering for  Robotics | Journal | JOSER | Software Engineering/ Robotics |
| [S45, S46] | Springer Tracts in Advanced Robotics | Book Chapter | STAR | Robotics |
| [S33, S35] | International Conference on  Simulation, Modeling, and Programming for Autonomous Robots | Conference | SIMPAR | Robotics |
| [S51, S55] | Workshop on Model-Driven Robot Software Engineering | Workshop | ***MORSE | Software Engineering |
| [S34, S38] | International Conference on Advanced Robotics | Conference | ICAR | Robotics |

*Prior to 2011 **SEAMS** was considered a workshop titled *Workshop on Software Engineering for Adaptive and Self-Managing Systems*. Now the title remains same but SEAMS is a symposium. The studies [S5, S36, S41] were all published before 2011 so we classify them as workshop papers.
**Prior to 2004 *IEEE Transactions on Automation Science and Engineering (**T-ASE**)* was titled as *IEEE Transactions on Robotics and Automation* with two included studies [S8, S28]. Now these two studies are organized under the new title.
***The first edition of this workshop was organized in 2014 promoting the application of software engineering and specifically the application of model-driven techniques to robotic systems

Among others, the main sources are *RAS, T-ASE*, *IROS* as some of the premier journals and conference series respectively for the robotic research community, whose studies are focused on supporting design and reconfiguration activities with a temporal distribution between 1998 to 2015 (cf. Figure 5). The top venues in software engineering community are *SEAMS* (a community of self-adaptation software research) and *ICSE* that represents the premier software engineering conference. The publications at these two venues are focused on runtime evolution (architectural reconfiguration) and design-time evolution (architectural reengineering) of robotic systems respectively with publication coverage from 2005 to 2009 with no publication identified at these two venues in the last five years. It is vital to mention about *STAR* with a specific book titled *Software Engineering for Experimental Robotics* [21] published in 2007 mainly addressing software engineering issues for robotic systems. The other two are *SIMPAR* and *ICAR* as robotics research venues with published studies focused on supporting robotics development and coordination related activities published between 2008 to 2012.

It is vital to mention that in recent years there is a focus on synergizing the research and practices between software engineering and robotics. A specific example is *JOSER* (first issue published in 2010) that aims at providing a platform where the existing software engineering approaches and methods can be leveraged for the development of robotic software systems. Moreover, *MORSE* (first edition in 2014) reflects an effort of software engineering community to support publishable and applicable results of model-driven engineering and its application to robotic systems. The sources and frequency of publication in Table 4 suggests that both the robotics and software engineering research communities are working to synergize their efforts by exploiting the models, languages, tools, infrastructures, patterns, principles and ecosystems etc. for the development, evolution and operations of robotic software. A recent special issue of Autonomous Robots reported the trends on the use of open source software for design and development of robotic systems [63].

## 7.3 Publications Distribution by Active Research Community

Now we report the findings aimed at answering RQ 2.3 for identifying the active communities of research on architecture-driven software for robotics. Figure 11 presents the relative distribution of the reviewed studies published by various communities. From Figure 11, it is clear that software architecture-driven robotics is multi-disciplinary research that ranges from *artificial intelligence* to *software*, *system and control and industrial engineering*. Figure 11 complements the findings presented in Table 4. The findings reveal that software engineering and robotics communities represent a combined 64% (30% and 34% respectively) of the reviewed studies. The distribution of the studies in terms of active research communities is based on the guidelines of ACM classification scheme and also on the analysis of the research focus of the publication venues where both communities have been publishing. For example, in Table 8 the study [S14] in Figure 11 is classified under computing and software engineering (publication source: *ICSE – International Conference on Software Engineering*). We have classified [S50, S54] under software engineering as *JOSER – Journal of Software Engineering for Robotics* (cf. Table 4) is a source of research that aims to support the application of software engineering methodologies to robotic systems.

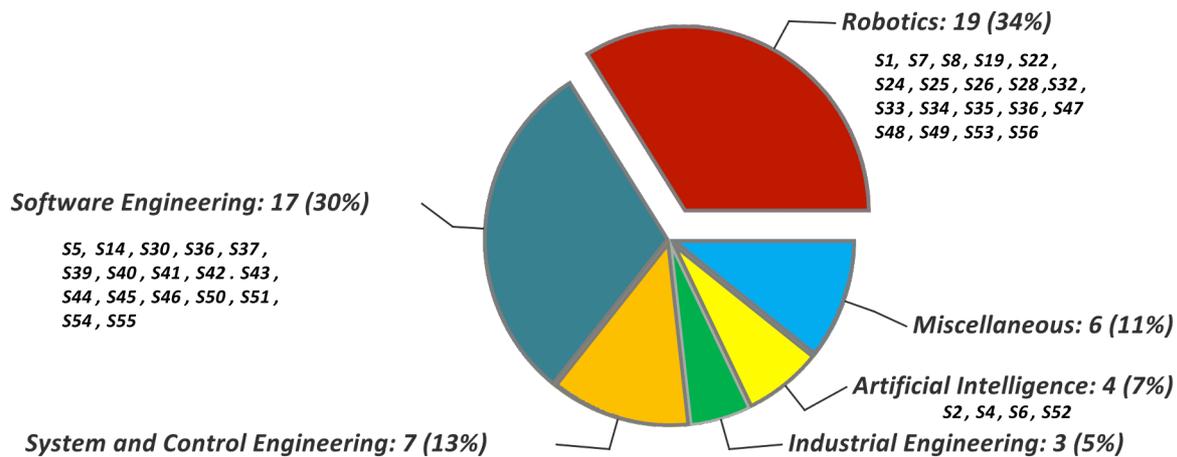

Figure 11. Overview of Study Distribution by Research Community

As a possible interpretation of Figure 11, the studies [S9, S12, S13] represent research on architecture for robotics in the context of *artificially intelligent systems to support* the coordination and development of robotics. One study [S9] exploits service-oriented robotics to develop a team of intelligent robots that collect and share mission specific information to support intelligent and autonomous first responder robots. The miscellaneous category refers to the studies published in other communities that are disjoint to the ones in Figure 11. For example, one study [S2] has been published in *IEEE Aerospace Conference* and proposes an architecture named CLARAty developed by NASA to support space control missions. The research demographics details in this section go beyond the mapping and classification of the existing research to present the publication frequency, sources and active communities. The demographic results suggest a multi-disciplinary research (cf. Figure 11) and possible collaborations among different research communities can further benefit the state-of-the-research. The results support the recent efforts that have been promoting cross-fertilization of research and practices from different communities to engineer robotic software [57, 58].

# 8. Discussion of Research Progression and Emerging Trends

Based on the taxonomical classification (RQ 1) and demographic information (RQ 2) of the reviewed research, we present our analysis of the research progression and the types of existing and emerging trends to answer RQ 3. We specifically aim to answer the sub-questions: *RQ 3.1: What are the past and active trends of research on architecting robotic software? and RQ 3.2: What emerging trends can be observed as an indication of future research?* First, we report the progress of the research (in Section 8.1). We then discuss some active trends of architectural solutions for robotics (in Section 8.2) answering RQ 3.1. We present some emerging trends as possible dimensions of future research (in Section 8.3) to answer RQ 3.2.

## 8.1 Research Progression for Software Architecture-Driven Robotics

The prominent trends of architecture-driven robotics research during (1999 – 2015) are shown in Figure 12. The research progress has been measured in terms of the advancement of solutions and trends that have emerged overtime. The research trend refers to a frequent solution as part of an identified research theme (cf. Section 4, Figure 4). For example, Figure 12 shows the trend of *programming models* or *frameworks* as solutions attributed to the high-level theme of *modeling, development and programming* of robotics using *OO-R* techniques. Figure 12 illustrates a transition of research trends in the context of OO-R, CB-R and SD-R that are referred to as architectural trends or architectural generations for robotic software that emerged and matured with more than two decades of research. In the context of Figure 12, an interesting observation is that the findings of the mapping study - progress and maturation of architectural solutions for robotic software - are consistent with the results of software technology maturation [24]. Specifically, in [24] a review of the growth and propagation of software technologies reveals that since its inception a software technology typically takes 15 - 20 years for wide-spread adoption and popularization. Figure 12 shows that early research on architectural solutions started in 1990s, however, the maturation and concentration of the state-of-the-research is only evident in the last decade (approximately 20 years later). We do not refer to or discuss the details of individual studies/solutions (as they represent specific cases). Instead we focus on higher-level themes. Based on the publication types and years (cf. RQ 2.1, Figure 10), we have identified three phases of the research progress. Considering Figure 12, no study from 1992 to 1998 has been included in our review – the reasons stated earlier (cf. Section 7.1).

- *Foundation Phase (1991 - 2004)* refers to the years of the earliest research on utilizing architectural models that primarily included the object-oriented modelling to enable the programming for robotic software in the context of OO-R. As illustrated in Figure 12, although there is less progress during this phase (number of studies and research themes), however, during this era of 15 years the concepts of architecture-driven robotics have been established. Specifically, architectural or design modeling is considered as the driver to develop reusable and modular programs. We have only identified 6 relevant studies (i.e., 11% approximately of the reviewed papers). After almost a decade through this phase during the early 2000, the research studies started to move beyond code-driven techniques to utilize the notion of component-based software engineering and specifically component-based robots to develop and evolve robotic software.

- *Maturation Phase (2005 - 2010)* represents a temporal view of research with consolidation and maturation of earlier solutions and the research progress aimed at establishing software architecture as an important artifact of robotic software development cycle. During this phase, a total of 30 studies (53%) have been identified that proposed innovative theory and solutions (such as modeling notations, and architectural frameworks) to extend or leverage the existing methods (robotics programming).

  One of the key advancements in this phase can be attributed to the application of component-based engineering techniques and specifically CB-R as a solution for reusability with off-the-shelf components. During this phase architectural modelling notations such as UML and ADLs (cf. Figure 7) became popular. The researchers also proposed or utilized various architectural frameworks (Figure 6) to support architecture-based development and evolution of robotic software.

– **Active Phase (2011 - 2015)** represents the recent progress in the area mostly based on CB-R and SD-R solutions. During this phase, the concept of service-orientation for robotic systems [18] has been extended to put forward the notion of cloud robotics [S20]. During this phase, a total of 20 studies (36%) have been published. The component-based techniques have been utilized to enable model-driven robotics. These trends have also been discussed later in detail as the possible dimensions of future research on this topic.

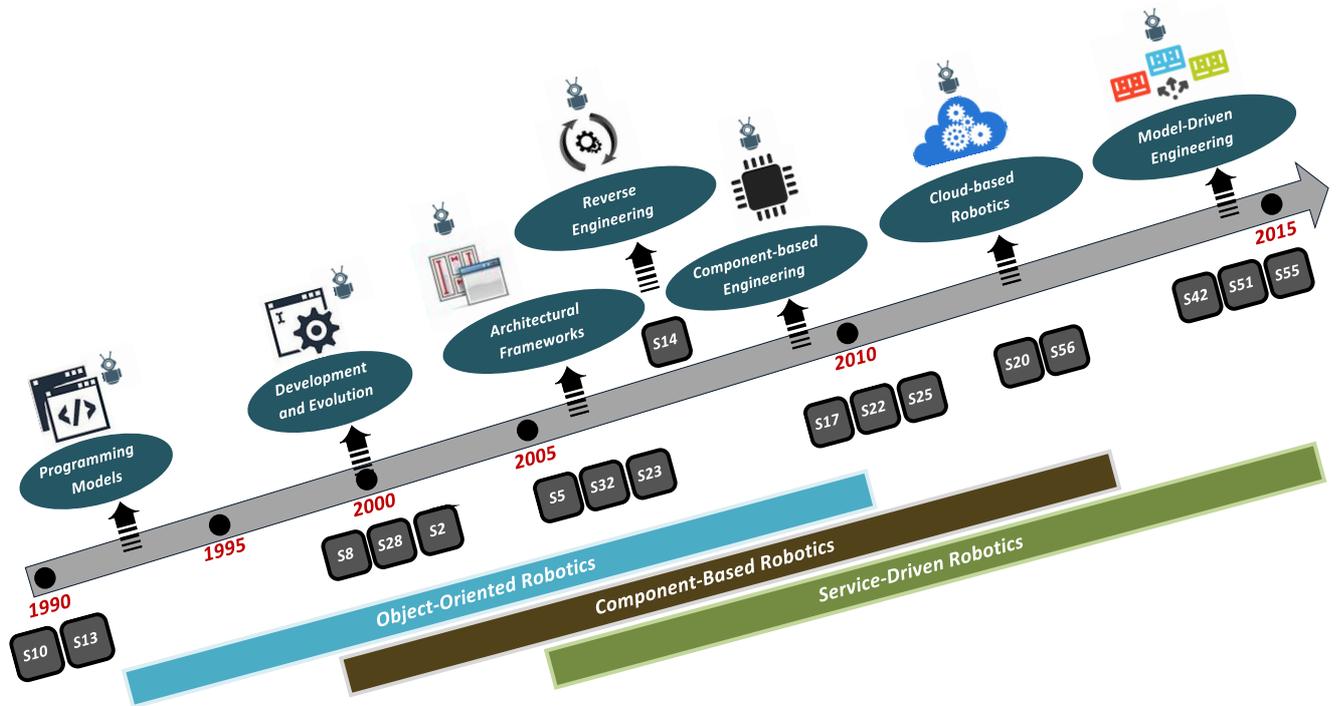

Figure 12. Overview of Research Progression with Emergent Trends.

## 8.2 Architectural Trends for Robotic Systems

An architectural trend (i.e., *architecture generation*) refers to *a direction of research in which architectural solutions are developed, matured and evolved overtime*. The discussion on this topic has been influenced by the findings about the present, past and future of software architecture [49]. We specifically highlight various architectural trends for robotics software based on the research progress in the area. In the context of architectural solution (cf. Section 5), we discuss the trends or generations of architectural solutions for robotics based on the commonality or distinction of the problems addressed and the solutions provided. We have identified three distinct trends: Object-oriented, Component-based and Service-driven robotics as presented in Table 5. Table 5 highlights the primary challenges these trends aim to address, the activity they support, summary of the solutions they offered and the relevant studies associated to them.

Our discussion is guided by Figure 13 that highlights a generic Reference Architecture (also presented in Section 2, Figure 1) that acts as a blueprint to map the individual trends or the specific architectural generation with a generic architecture common to all. An architectural overview of each trend has been provided to highlight how architectural solutions in individual trends bridge the gap between robotics hardware and system software.

**Trend I - Object-Oriented Robotics (OO-R)**

*Overview* - OO-R is the first research trend that exploits the principle of object-orientation to provide a modular and reusable objects and frameworks to support the programming of robotics – i.e., support for development activities. OO-R represents some of the earliest evidence (published in early 90's, 1991 [S10] and 1992[S13]) on design and architectural support for robotics development. In terms of representation, the view of the OO-R has been provided in Figure 13 A). The view represents a generic and consolidated representation based on the analysis of architectural solutions earlier – individual solutions may have more solution-specific details. The generic architecture for OO-R consists of three distinct elements namely: *Core*, *Generic* and *Specific* along with two types of interconnections of elements named: *uses* and *inherits* (common OO relations). The hierarchy specifies that structural overview that generic components must have core components, while specific components must inherit from generic components. The hierarchal structure for OO-R cannot be altered, for example, specific components must rely on a mediator (generic components) to access the core. An upside view of Figure 13 A), OO-R is similar to the layered architecture [42], where each layer encapsulates specific/partial functionality and relies on a layer directly below or above it.

*Support for Robotics* – Based on OO-R representation in Figure 13 A), to support the development of a robot, the core component encapsulates the low-level control and manipulation logic of a robot in terms of the hardware drivers and configurations that enable software interaction(s) with the hardware. Instead of developing same logic each time for different systems, the generic components must utilize the core to provide the general functionality. A specific component can inherit the functionality from generic ones to support a fully functional robot with customized functionality; for example, the case of a robotic arm in [S13]:

  *Step I* - Core provides the necessary control logic to access robotic arm through hardware driver,

  *Step II* - Generic layer can reuse the control logic to enable the movement of robotic arm, and finally

  *Step III* - Specific layer extends the generic functionality to rotate the arm at a specific angle and precision.

The studies incorporating OO-R [S10, S13] represent one of the earliest efforts in exploiting design and architectural specification to abstract the low-level programming details (through Core component) with modular and reusable components (i.e., Generic and Specific). These approaches led to the development of framework(s) and language(s) such as RIPL (Robot Independent Programming Language) [S13] and Robot Markup Language (RML) [S12] to support OO-R systems [S15, S20]. *Architectural notations* for OO-R are modeled with UML diagrams such as *class* and *sequence* diagrams (cf. Figure 7).

Table 5. Overview of the Architectural Trends for Robotics

| Trend / Challenges | Objected-Oriented Robotics (OO-R) | Component-Based Robotics (CB-R) | Service-Driven Robotics (SD-R) |
|---|---|---|---|
| Reduced Complexity | ✔ | ✔ | ✔ |
| Increased Modularity | ✔ | ✔ | ✔ |
| Software Reusability | ✔ | ✔ | ✔ |
| Model-based Development | | ✔ | ✔ |
| Software Distribution | | | ✔ |
| Software Composition | | ✔ | ✔ |
| **Activity** | | | |
| Development | ✔ | ✔ | ✔ |
| Operations | | | ✔ |
| Evolution | | ✔ | |
| **Solution** | Objects to model and encapsulate reusable source-code for robotics. | Off-the-shelf architectural components to model and develop robotics. | Loosely-coupled, dynamically composable services for robotics. |
| **Studies** | S10, S12, S13, S18, S21 | S17, S22, S23, S24, S25, S28, S29, S30, S31, S32, S33, S34, S35, S36, S37, S38, S39, S40, S41, S42, S43, S44, S45, S46, S48, S49, S50, S53, S54 | S6, S7, S9, S15, S16, S19, S26, S27, S55, S56 |
| The studies [S14, S20] could not be associated with any of three trends as: [S14] is an exception as a single study on architecture-based re-engineering that does not relate to any of the trends. [S20] discusses cloud-based robots (with only implicit reference to software services for developing cloud-robots). | | | |

## Trend II - Component-based Robotics (CB-R)

*Overview* - CB-R follows the component-based software engineering techniques. It utilizes architectural components to support the development, evolution and operations of robotics [27]. CB-R also represents some of the earlier and most recent solutions (covering studies from 1998 [S8] to 2013 [S40]) on exploiting architectural component for robotic systems.

The structural view of CB-R in Figure 13 B) is based on component composition and component interaction [27]. OO-R utilized three distinct elements namely: *Provider*, *Requester* and Component *Repository*. The Provider and Requester (as a composite) elements are connected using *request* and *provide* interconnections. The requester element requests the functionality provided by the provider, while the component repository provides a collection of reusable and off-the-shelf components [13, 21]. The CB-R, aims to abstract the complexity of programming by means of reusable and off-the-shelf components. There are some specialized versions of component-based robotics that are developed by utilizing some well-known design/architectural patterns and techniques such as *client-server* [S3, S31, S22], *layered* [S5, S23, S24] and *model-driven* [S33, S38, S40] architecture.

*Support for Robotics:* Based on CB-R representation in Figure 13 B), to support robotics the control component abstracts the low-level control and manipulation logic of a robot. The requester component is a composite element based on its child components each of which can be responsible for a specific functionality to complete the functionality of its parents. For example, the requester element represents navigation (composite functionality) supported by combining partial functionalities such as localization, mapping and movement (atomic functionality). The control component acts as a provider, providing the necessary drivers and information to customize the robotics navigation. The requester component can utilize this functionality with its child component to support the robotic navigation in a leader-follower scenario. The component repository plays an important role in a way that it provides a collection of reusable components, also called off-the-shelf components that can be used for development. As presented in Figure 13 to highlight a difference between OO-R and CB-R is that, OO-R supports object reusability (at source code level) with inheritance type relationship in Figure 13 A). In contrast, component reusability (at architecture level) is supported with component composition in Figure 13 B). CB-R can rely on model-driven engineering techniques for an automatic generation of code by higher-level components. The CB-R techniques mostly utilized the UML models as well as ADLs for architectural representation and description.

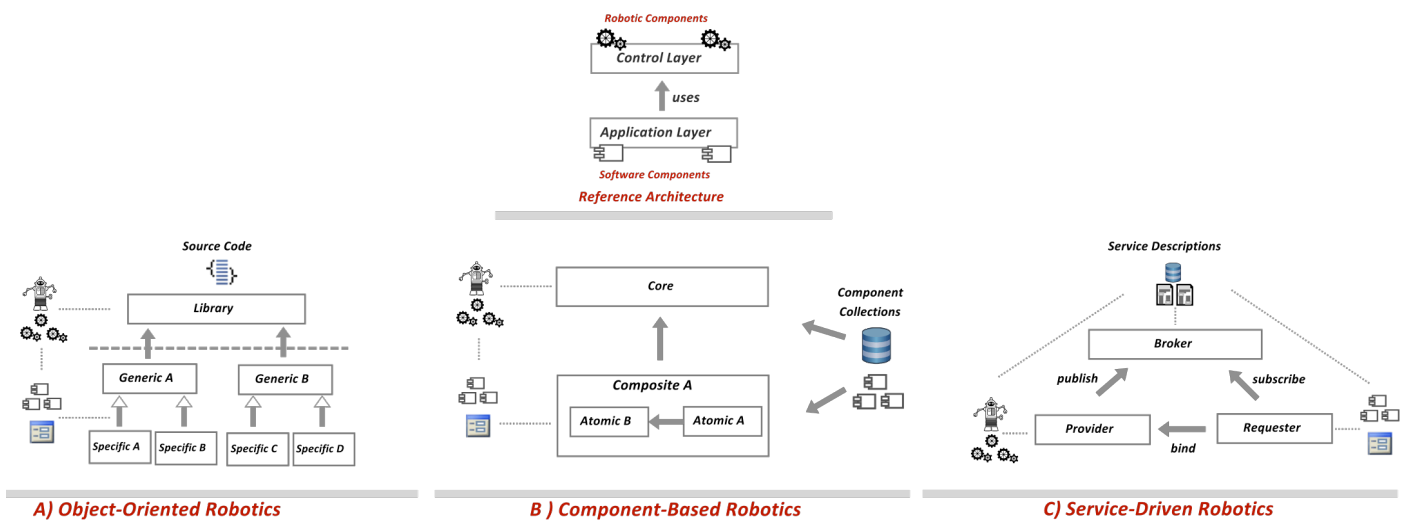

Figure 13. An Overview of Various Architectural Trends

**Trend III - Service-Driven Robotics (SD-R)**

*Overview* - The SD-R is a representation of service-driven software or specifically Service Oriented Architecture (SOA) that relies on loosely coupled and dynamically composed software services for robotics. SD-R represents the recent and emerging solutions (published in years: 2012 [S20], 2013 [S26], 2014 [S55, S56]). Architectural view of the SD-R has been provided in Figure 13 C) that is based on the classical

Service-Oriented Architecture (SOA) framework or other well-known patterns (mediator, publish-subscribe [43]). The architectural representation for SD-R consists of elements named: *Broker*, *Publisher* and *Subscriber* of services along with their interconnections namely *provide*, *request* and *bind*. The constraints on the structure or the role of each component are specified based on the order of their interactions, such that (Step I) *Publisher* provides a description of service to the *Broker*, (Step II) *Subscriber* subscribes to the *Broker* for a notification of available services, and finally (Step III) *Subscriber* binds to *Subscriber* for required services - (Step I) and (Step II) above can be interchanged.

*Support for Robotics* - Based on SD-R view in Figure 13 C), the publisher is a component responsible for providing software services that enable the interaction with hardware robotics. The services include the control or manipulation logic of the robotics hardware. There are two scenarios, where (i) these services are provided along with hardware components [S27] or (ii) can be developed separately [S15], in both cases their descriptions (not implementation) need to be published to the Broker. The description can include the name, description of the provided services in terms of service interface using Web Service Description Language. Broker is simply a service description repository or directory that mediates between publisher and subscriber. When a description matches the requirements of the subscriber the publisher can be bound to offer services to the subscriber. The studies incorporating SD-R such as [S26, S27] are more recent trends to exploit services for robotics. The service-oriented robots have found their application in home service [S6, S7] and mission-critical robots [S9, S26]. An interesting study [S27] has focused on developing robot as a service as a notion of cloud-based robotics. The prominent architecture notations for SD-R are UML *state-chart* [S6, S7], *deployment* [S9], *component* [15, 26, 27] and *class diagram* [S16].

## 8.3 Emerging Trends for Software Architecture-Driven Robotics

Based on an overview of the research progress (Figure 12) and the extended details of the above mentioned phases, we now discuss the emerging trends as dimensions of future research to answer RQ 3.2.

– **Model-Driven Engineering of Robotic Software** aims at developing models by utilizing the model transformation techniques for supporting robotic software*.* Component-based software engineering [21] or specifically CB-R [8, 13] aims to minimize the inherent complexity associated with programming and enhance the reusability with architectural components for robotic software. Despite the benefits of CB-R (Section 8.2), architectural components are an abstraction of implementation-specific details and must be refined to derive the executable code. In this context, model-driven engineering [S38] provides a solution by capturing the overall system model (with components) that is refined and transformed into executable model (with source code). Figure 12 indicates that in recent years there is a growing interest in applying model-driven solutions to robotic software[7]. The model-driven engineering for robotics provides a methodology – exploiting languages, patterns, tools and infrastructures - to support the development, evolution and operation of robotic applications. It can be argued that CB-R has paved the way for model-driven robotics as architectural components bridge the gap between the requirements of a specific domain (navigation [40], automation [33], Figure 9) and satisfying those requirements with executable code. Model-driven robotics relies on three levels of abstraction:

- *Meta-model Level capturing of Domain Assumptions and Requirements*: first of all, the requirements or the assumptions of the domain are captured and represented as a meta-model that drives model-driven development. For example, the domain requirements may enforce team communication among robots using a central coordinator (mediator) rather than a robot-to-robot interaction (peer to peer).

---

[7] We observed such research interests in some of most recent studies and efforts such as the *Workshop on Model-Driven Robot Software Engineering* to organize a community for promoting research and practices specific to the application of model-driven solutions to robotic systems.

- *Platform Independent Level representing Architectural Components*: once the requirements have been captured, they are translated into components for architectural modeling independent of any (execution or deployment) platform. For example, based on the requirement specified above a broker component needs to be integrated between two robots for their coordination.

- *Platform Specific Level generating Executable Code*: finally the generic architectural components are transformed into executable source code for a specific execution platform.

The studies exploited model-driven techniques for the development and runtime evolution of robotic systems. In contrast to OO-R and CB-R, with model-driven robotics there is a greater emphasis and efforts on creating models. However, once models are established they help with minimizing the programming complexity (from OO-R) and support reusability across various platforms (compared to the conventional CB-R).

- **Cloud Robotics** appears to be the next generation trend for robotic systems, which are expected to exploit the principle of service-orientation to develop and provide robotic software services. Such software services can be dynamically discovered, composed and executed to support SD-R [18, S55, S56] (cf. Section 8.2). The emergence of cloud robotics is a result of service-orientation and more specifically the cloud computing model [56]. In a cloud computing model, instead of an upfront acquisition or deployment of computing resources, cloud resources can be utilized as pay-per-use services for hardware and software resources that are publically available [51]. Cloud robotics enables the assembly or operations of a robot by utilizing the powerful computation, virtually unlimited storage as well as communication resources available from cloud-based infrastructures. This means a robot can discover and utilize distributed hardware and software components on the Internet. Cloud computing is specifically beneficial for developing mobile robots [55] in which on-board computation can be offloaded to processors and data stores residing in a cloud. RobotEarth[8] is a relevant project that is focused on the concept to create a World Wide Web for robots where robots can upload/download, and share the knowledge about their operational environment. This creates a web of co-operative robots that are autonomous individually and can also work as a team for an efficient completion of designated tasks. In the following, we clarify the similarities and distinctions among *service* and *cloud-based* architectures for robotic software

- **Cloud vs Service-based Architectures for Robotic Software** in the context of software architecture, the terms *service* and *cloud* computing are complementary, both denoting the architectural style(s) and enabling technologies to develop and deploy architectural components as dynamically composed software services [11, 15]. However, we must distinguish between the concept of service-driven [14] and cloud-based robotics [15] while also acknowledging the fact that (software) services remain fundamental to both types. To maintain the distinction, we first look at the established definitions of both the service and cloud-based architectures and then exemplify them each with the help of Figure 14. According to the definition by Open Group a *service-driven* or *service-oriented architecture* (SOA) is an architectural style that exploits loosely coupled - heterogeneous and dynamically composed - *software services* that can be *published* (by service provider) via a *broker*, *discovered* and *subscribed* (by service requester) to develop and deploy distributed system as illustrated in Figure 14 a). The service broker acts as a mediator between the service providers and requesters. For example a team of first responder robots that coordinate and navigate together [14] rely on dynamically discovered and composed software services to accomplish their mission. Specifically, each of the robotic functionality such as obstacle avoidance, team-lead following or searching the target is supported in form of software services provided by the mission control server. In contrast to the built-in software functionality (pre-packaged modules/components) for traditional robots, the service-driven robot (service requester) can communicate with the central server (service provider via broker) to acquire the required functionality by exploiting software services, whenever required.

---

[8] RoboEarth: http://roboearth.org/cloud_robotics/

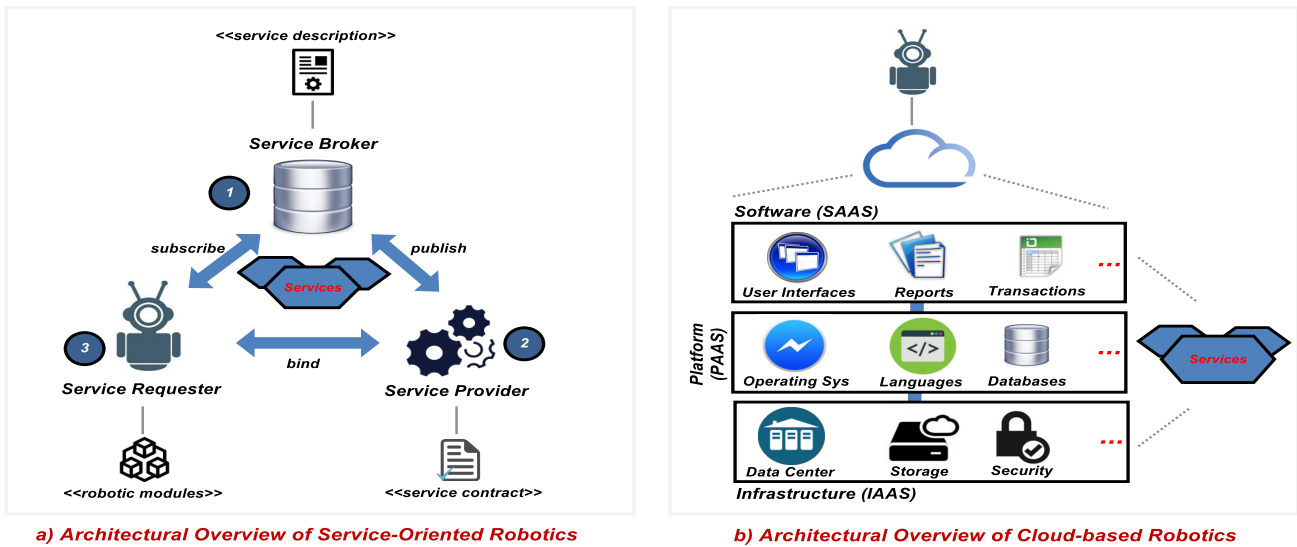

a) Architectural Overview of Service-Oriented Robotics          b) Architectural Overview of Cloud-based Robotics

Figure 14 An Illustrative Overview of Service and Cloud-Based Architectures for Robotic Software

NIST defines *cloud computing* as a combination of *software*, *platform* and *infrastructure* services that enable the entities or organizations to leverage distributed and interoperable services to develop and deploy their software systems over available resources (such as public or private servers), represented in Figure 14 b). In comparison to the service provisioning as in SOA, the cloud computing model also offers platform and necessary infrastructure (to service requesters) for off-loading the computation and memory-intensive tasks. For example, in a robot that relies on Simultaneous Localization And Mapping (SLAM) operations, the computationally expensive map optimization and storage tasks are off-loaded to a Cloud [S56], whereas tasks like camera tracking is performed by the robot itself. In this scenario, a robot's onboard computers are freed from most of computation, the only extra requirement is network connectivity.

We conclude that software services or specifically SOA is fundamental to cloud-based architectures. SOA focuses on a collection of services to bind the service providers and requesters in a distributed architecture. In comparison, a cloud-based architecture goes beyond software services (e.g., user interfaces and transactions) to offer platforms (e.g., operating systems and databases) and infrastructure (e.g., storage centre and security servers). In comparison to SOA that relies on traditional means of software quality such as modularity, heterogeneity, extensibility, a cloud-based architectures must also satisfy quality attributes like elasticity, multi-tenancy and resource virtualization [15, 51].

Considering the context of cloud robotics, it is vital to distinguish between **cloud-native** and **cloud-enabled** robotics [51]. Cloud-native are the types of robots that are specifically built to operate in cloud-based environment [S20, S27, S56]. In contrast, the cloud-enabled robots refer to the existing or legacy robots that need to be evolved or customized to enable them exploit the cloud computing model. Therefore, evolving a legacy robot towards a cloud-enabled robot requires additional efforts; however, such an evolution can enable a robot system to exploit cloud-based resources. There is a lot of progress on research that enables the migration or evolution of the legacy software systems to cloud-enabled software [52] and the existing research can be influential to support robotics evolution. In this context, the study [S14] highlights a solution on architectural reengineering to support the evolution of legacy architecture to a new architecture. Architectural reengineering solutions and specifically the models of legacy migration towards cloud [52] can facilitate *architecture-driven migration of legacy robots to cloud-enabled robotics*.

# 9. Validity Threats

This study provides a classification to map the reported solutions by reviewing and analyzing peer-reviewed literature. We followed the guidelines for conducting systematic mapping studies reported in [16, 17, 50] based on a defined and - internally and externally - evaluated protocol for SMS (cf. Section 3). Like any other empirical study, systematic mapping studies can also have limitations that must be considered for analyzing the potential impact of the validity threats to the findings of SMS [50]. We discuss three types of validity threats associated with different activities of this SMS.

– ***Threats to Identification of Primary Studies***. In the literature search strategy (cf. Section 3.2), we aimed to retrieve as many relevant studies as possible to avoid any possible literature selection bias and to accommodate all the available evidence. We faced a challenge in determining the scope of our study as the notion of architecture means different things to different research communities including software engineering, robotics, artificial intelligence and others (cf. Figure 11). Therefore, to cover them all and avoid any bias, we searched the literature based on relevant terms (cf. Table 2) and combined them in our search string (cf. Figure 3). While this search strategy and search string composition significantly increases the search work [45], however, it enabled us to find a comprehensive set of the relevant study. We also developed and evaluated a review protocol (cf. Section 3.1). The protocol provides a replicable blue-print to derive the search strategies, literature identification and selection.

– ***Threats to Quality of Studies and Data Extraction Consistency.*** The results of this study and their quality are based on the quality of the studies that have been reviewed. This means if the quality of the primary studies is low, the claims and their supporting evidence derived from these studies are unlikely to be strong and reliable. Therefore, it is vital to (i) minimize the threats regarding the quality of selected studies and to ensure (ii) a consistent representation of data extracted from these studies. It is of central importance to qualitatively analyze and synthesize of the data extracted from the selected studies [50]. As described in section 3, we followed a multi-step process and explicitly assessed the quality of each individual study to ensure that a lack of quality results in an exclusion of the study. The ideal scenarios may strictly adhere to the guidelines in [16, 50], however, the quality metric (cf. Appendix B.2, Table II, Table III) can be subjective based on the objectives of SMS and the consensus among researchers.

Moreover, we derived a structured template (cf. Table IV) to ensure consistency in data extraction and capturing as per the needs of the study's RQs. For a fine-grained representation of the extracted data, we have defined a generic and mapping specific attributes to capture data for detailed synthesis.

– ***Threats to Data synthesis and Results Reporting.*** The final type of threat corresponds to the bias or a lack of systematic approach to synthesize and report the results. We tried to mitigate this threat by conducting a pilot study. A limited number of researchers and their expertise (software engineering and software architecture) may have an internal bias on the style and reporting of results. We tried to minimize this by (i) classifying the studies (cf. Figure 4) by following the ACM classification scheme for computing research, guidelines from existing studies [44], our experience from [38], (ii) mapping the results (cf. Figure 5) to better organize the overlapping and disjoint findings and (iii) an external evaluation of the review protocol.

The threat to the reliability of data synthesis and reporting has been mitigated based on discussion and peer review of the extracted data by the researchers, having a structured template for data synthesis, and several steps where the scheme and process were refined and evaluated. Whilst we followed the guidelines from [17, 50] to conduct the study, we had deviations from the ideal approaches based on the requirements of this study detailed in Section 3. We believe that the validity of the study is high, given the use of a systematic and recommended procedure, the extensive discussions and evaluation of the protocol and a pilot study to refine the scope of review.

# 10. Conclusions

The goal of this SMS was to systematically identify, select, and analyze the published research that reports architecture related solutions for developing and evolving robotic systems. This SMS has identified and discussed various architecture-related challenges and solutions reported for robotic systems. The results of the classification and mapping of the existing research have been provided in terms of structured tables and illustrative figures with the aim to systematized and disseminate the knowledge about architectural challenges and solutions for robotic systems. The taxonomical classification (cf. Figure 4, Section 4) provides a holistic overview of the overlapping and distinct research themes and their sub-themes that emerged and progressed over time (early 90s – to 2015). The study complements the existing research in terms of the systematic reviews [18, 19] and surveys-based studies [53, 54, 55] on the application of various software engineering techniques to robotic systems. The key contribution of this research is to specifically investigate the state-of-research on the role of software architecture in robotic systems.

The classification and mapping along with their accompanying templates (cf. Table IV, Appendix B) provides 10 data items as a moderate amount of extracted information from each of the reviewed studies. We systematically reviewed 56 studies using 10 data items that resulted in a collection of 56 x 10 = 560 potentially unique elements of investigation. Based on this, considering an example case [S39], a reader can analyze the information as:

– **Subject**: analyze *Architecture-driven Adaptation (problem space)*
– **Object**: using *Model-driven Engineering (solution space)*
– **Implications:** for *mission critical robots*.

***Architecture-based Engineering for Robotic Software*** – In recent years, multi-disciplinary efforts have been promoting cross-fertilization of existing research and practice to model, develop and evolve complex robotic systems (cf. Section 7.3, RQ 2.3). Our study aims to support some emerging efforts focused on synergizing the current and future research on Robotics Systems and Software Engineering [57, 58]. Specifically, the mapping study is expected to highlight the role of software architecture for the development and evolution of large-scale robotic systems. The outcomes of our study highlight some innovative and frequently used architectural solutions to address various challenges of developing and evolving robotic systems. The study findings also provide inspiration and ideas for future research that is expected to contribute to the development of the next generation of robotic systems as part of multi-disciplinary research and development efforts. The results from this mapping study can have several implications for software architecture and robotic systems researchers and practitioners:

– *Researchers* who need to quickly access a body of knowledge based on relevant literature and solutions supporting the role of software architecture for robotics. The study also highlights features of existing and emerging dimensions of the research. Moreover, a mapping (cf. Figure 5, Section 4) of the research state-of-the-art and its historical progression can help identify aspects like: (i) available evidence of research progression, (ii) temporal distribution and frequency of various research themes, (iii) the focus and contributions of most and the least frequent themes, etc. Table 2 can be interpreted as a structured catalogue – mapping challenges and solutions – to help us quickly identify various architectural solutions and the role of software architecture in robotic systems.

– *Practitioners* interested in understanding the academic research in terms of existing solutions and frameworks (academic, industrial, and open-source) along with modeling notations, evaluation methodologies to analyze possibilities where academic approaches can be leveraged for industrial solutions. The discussion of architectural solutions has been complemented with presentation of some relevant frameworks (cf. Figure 4, Section 4) that support these solutions. The interpretations can be diverse but the aim of this study is to provide a foundation to consolidate, analyze and present the impact of the existing research on architecture-driven solutions for robotic systems. The mapping study also highlights the needs for the future research on analyzing practitioners' views on the role of software architecture for robotic systems that is currently lacking.

We conclude the reporting of this SMS by highlighting two core findings of this study:

– A systematic analysis of more than two decades of research reveals three distinct architectural trends or generations that emerged and referred to as OO-R, CB-R and SD-R in order of their emergence. In the last decade, there has been a noticeable growth of research with various innovative solutions, however, model-driven and cloud-based robotics appear to be the most prominent and emerging trends. Model-driven robotics bridges the gap between domain requirements and executable specifications, whereas cloud robotics exploits the concepts of SD-R to construct a web of distributed and connected robots.

– The conventional concepts of architectural frameworks and architectural notations are well integrated into the robotics domain. UML-based modeling is influential and utilized by researchers and practitioners (other than software architecture or software engineering domain) to model, develop and evolve software architecture for robotics. Researchers have used various validation methods to evaluate the reported solutions; however, there is a clear lack of focus on architecture specific evaluation for validating the quality-oriented or architecturally significant requirements for robotic software.

# Acknowledgements

The authors would like to thank Dr. Pooyan Jamshidi (affiliated with: Lero – the Irish Software Engineering Research Centre, Dublin City University, Ireland) for his feedback and thoughtful suggestions throughout the development and evaluation of the review protocol.

# Appendix A

## List of Selected Studies for Systematic Mapping

| Study ID | Author(s), Title, Channel of publication | Publication Year | Citation Count | Quality Score |
|---|---|---|---|---|
| [S1] | T. Kaupp, A. Brooks, B. Upcroft and A. Makarenko. *Building a Software Architecture for a Human-Robot Team Using the Orca Framework*. In IEEE International Conference on Robotics and Automation | 2007 | 07 | 3.5 |
| [S2] | R. Volpe, I.A.D. Nesnas, T. Estlin, D. Mutz, R. Petras, H. Das. *The CLARAty Architecture for Robotic Autonomy*. In IEEE Aerospace Conference | 2001 | 232 | 2.5 |
| [S3] | T. Baier, M. Hüser, D. Westhoff, J. Zhang. *A Flexible Software Architecture for Multi-modal Service Robots*. In Multiconference on Computational Engineering in Systems Applications | 2006 | 09 | 2.5 |
| [S4] | H. Ahn, D-S. Lee, S. C. Ahn. *A Hierarchical Fault Tolerant Architecture For Component-based Service Robots*. In IEEE International Conference on Industrial Informatics. | 2010 | 06 | 2.5 |
| [S5] | G. Edwards, J. Garcia, H. Tajalli, D. Popescu, N. Medvidovic, G. Sukhatme. *Architecture-driven Self-adaptation and Self-management in Robotics Systems*. In ICSE Workshop on Software Engineering for Adaptive and Self-Managing Systems. | 2009 | 32 | 3.0 |
| [S6] | X. Ma, K. Qian, X. Dai, F. Fang, Y. Xing. *Framework Design for Distributed Service Robotic Systems*. 5th IEEE Conference on Industrial Electronics and Applications. | 2010 | 0 | 2.5 |
| [S7] | T. Maenad, A. Tikanmäki , J. Riekki , J. Röning. *A Distributed Architecture for Executing Complex Tasks with Multiple Robots* In IEEE International Conference on Robotics and Automation | 2004 | 16 | 3.0 |
| [S8] | L. E. Parker. *ALLIANCE: An Architecture for Fault Tolerant Multirobot Cooperation*. In IEEE Transactions on Robotics and Automation | 1998 | 1206 | 4.5 |
| [S9] | J. S. Cepeda, R. Soto, J. L. Gordillo, L. Chaimowicz. *Towards a Service-Oriented Architecture for Teams of Heterogeneous Autonomous Robots*. In 10th Mexican International Conference on Artificial Intelligence | 2011 | 01 | 3.0 |
| [S10] | D. J. Miller, R. C. Lennox. *An Object-oriented Environment for Robot System Architectures*. In IEEE Control Systems | 1991 | 136 | 4.0 |
| [S11] | J. F. Inglés-Romero, C. Vicente-Chicote, B. Morin, O. Barais. *Towards the Automatic Generation of Self-Adaptive Robotics Software: An Experience Report*. In 20th IEEE International Workshops on Enabling Technologies: Infrastructure for Collaborative Enterprises. | 2011 | 03 | 2.5 |
| [S12] | J. Kwak, J. Y. Yoon, and R. H. Shinn. *An Intelligent Robot Architecture based on Robot Mark-up Languages*. In IEEE International Conference on Engineering of Intelligent Systems. | 2006 | 07 | 2.5 |
| [S13] | P. Kazanzides, J. Zuhars, B. Mittelstadt, B. Williamson, P. Cain, F. Smith, L. Rose, B. Musits. *Architecture of a Surgical Robot*. In IEEE International Conference on Systems, Man and Cybernetics. | 1992 | 19 | 3.0 |
| [S14] | M. Kim, J. Lee, K. C. Kang, Y. Hong, S. Bang. *Re-engineering Software Architecture of Home Service Robots: A Case Study*. In 27th International Conference on Software Engineering. | 2005 | 16 | 4.0 |
| [S15] | J. Yool, S. Kim, and S. Hong. *The Robot Software Communications Architecture (RSCA): QoS-Aware Middleware for Networked Service Robots*. In SICE-ICASE International Joint Conference | 2006 | 11 | 3.0 |
| [S16] | W. Hongxing, L. Shiyi, Z. Ying, Y. Liang, W. Tianmiao. *A Middleware Based Control Architecture for Modular Robot Systems*. In IEEE/ASME International Conference on Mechtronic and Embedded Systems and Applications. | 2008 | 09 | 2.5 |
| [S17] | N. Ando, T. Suehiro, K. Kitagaki, T. Kotoku and W. Yoon. *Composite Component Framework for RT-Middleware (Robot Technology Middleware)*. In IEEE/ASME International Conference on Advanced Intelligent Mechatronics. | 2005 | 18 | 3.0 |
| [S18] | I. Buzurovic, T. K. Podder, L. Fu, and Y. Yu. *Modular Software Design for Brachytherapy Image-Guided Robotic Systems*. In 2010 IEEE International Conference on Bioinformatics and Bioengineering. | 2010 | 0 | 3.0 |
| [S19] | S. Limsoonthrakul, M. N. Dailey, M. Srisupundit. *A Modular System Architecture for Autonomous Robots Based on Blackboard and Publish-Subscribe Mechanisms*. In International Conference on Robotics and Biomimetics. | 2008 | 09 | 2.5 |
| [S20] | G. Hu, W. P. Tay, and Y. Wen. *Cloud Robotics: Architecture, Challenges and Applications*. In IEEE Network. | 2012 | 30 | 3.5 |
| [S21] | A. Angerer, A. Hoffmann, F. Ortmeier, M. Vistein and W. Reif. *Object-Centric Programming: A New Modeling Paradigm for Robotic Applications*. In International Conference on Automation and Logistics | 2009 | 01 | 2.5 |
| [S22] | M. Y. Jung, A. Deguet, and P. Kazanzides. *A Component-based Architecture for Flexible Integration of Robotic Systems*. In IEEE/RSJ International Conference on Intelligent Robots and Systems. | 2010 | 19 | 2.5 |
| [S23] | D. Kim and S. Park.  *Designing Dynamic Software Architecture for Home Service Robot Software*. In International Conference on Embedded and Ubiquitous Computing. | 2007 | 05 | 4.0 |
| [S24] | Y. Park, I. Ko, and S. Park. *A Task-based Approach to Generate Optimal Software-Architecture for Intelligent Service Robots*. 16th IEEE International Conference on Robot & Human Interactive | 2007 | 01 | 2.5 |
| | Communication. | | | |
| [S25] | W. Hongxing, D. Xinming, L. Shiyi, T. Guofeng, W. Tianmiao. *A Component Based Design Framework for Robot Software Architecture*. IEEE/RSJ International Conference on Intelligent Robots and Systems. | 2009 | 06 | **2.5** |
| [S26] | Lorenzo Flückiger, Hans Utz, *Service Oriented Robotic Architecture for Space Robotics: Design, Testing, and Lessons Learned*. Journal of Field Robotics | 2013 | 01 | **4.0** |
| [S27] | Anis Koubaa, *A Service-Oriented Architecture for Virtualizing Robots in Robot-as-a-Service Clouds*. 27th International Conference on Architecture of Computing Systems | 2014 | 0 | **3.0** |
| [S28] | James E. Beck, Michael Reagin, Thomas E. Sweeny, Ronald L. Anderson, Timothy D. Garner. *Applying a Component-Based Software Architecture to Robotic Workcell Applications.* In IEEE Transactions on Robotics and Automation. | 2000 | 17 | **4.0** |
| [S29] | Javier Gamez García, Juan Gómez Ortega, Alejandro Sánchez García, and Silvia Satorres Martínez. *Robotic Software Architecture for Multisensor Fusion System.* IEEE Transactions on Industrial Electronics | 2009 | 32 | **3.5** |
| [S30] | Nenad Medvidovic, Hossein Tajalli, Joshua Garcia, and Ivo Krka, Yuriy Brun, George Edwards. *Engineering Heterogeneous Robotics Systems: A Software Architecture-Based Approach.* IEEE Computer. | 2011 | 07 | **4.0** |
| [S31] | Soohee Han, Mi-sook Kim, and Hong Seong Park. *Development of an Open Software Platform for Robotic Services.* IEEE Transactions on Automation Science and Engineering | 2012 | 12 | **4.0** |
| [S32] | Alex Brooks, Tobias Kaupp, Alexei Makarenko and Stefan Williams, Anders Oreback. *Towards Component-based Robotics.* IEEE/RSJ International Conference on Intelligent Robots and Systems | 2005 | 168 | **2.5** |
| [S33] | Li Hsien Yoong, Zeeshan E. Bhatti, and Partha S. Roop. *Combining IEC 61499 Model-Based Design with Component-Based Architecture for Robotics.* The Third International Conference on Simulation, Modeling, and Programming for Autonomous Robots | 2012 | 01 | **2.5** |
| [S34] | Jonghoon Kim, Mun-Taek Choi, Munsang Kim, Suntae Kim, Minseong Kim, Sooyong Park, Jaeho Lee, ByungKook Kim. *Intelligent Robot Software Architecture.* 13th International Conference on Advanced Robotics. | 2008 | 07 | **2.5** |
| [S35] | Jonghoon Kim, Mun-Taek Choi, Munsang Kim, Suntae Kim, Minseong Kim, Sooyong Park, Jaeho Lee, ByungKook Kim. *Intelligent Robot Software Architecture.* 13th International Conference on Advanced Robotics. | 2008 | 07 | **2.5** |
| [S36] | Dongsun Kim, Sooyong Park, Youngkyun Jin, Hyeongsoo Chang, Yu-Sik Park, In-Young Ko, Kwanwoo Lee, Junhee Lee, Yeon-Chool Park, Sukhan Lee. *SHAGE: A Framework for self-Managed Robot Software.* In 2006 International Workshop on Self-adaptation and Self-managing Systems | 2006 | 22 | **2.5** |
| [S37] | Minseong Kim, Suntae Kim, Sooyong Park, Mun-Taek Choi, Munsang Kim, Hassan Gomaa. *UML-based Service Robot Software Development: A Case Study.* 28th International Conference on Software Engineering. | 2006 | 29 | **3.0** |
| [S38] | Christian Schlegel, Thomas Haßler, Alex Lotz and Andreas Steck. *Robotic Software Systems: From Code-Driven to Model-Driven Designs.* International Conference on Advanced Robotics | 2009 | 39 | **3.0** |
| [S39] | Hossein Tajalli, Joshua Garcia, George Edwards and Nenad Medvidovic. *PLASMA: A Plan-based Layered Architecture for Software Model-driven Adaptation.* IEEE/ACM International Conference on Automated Software Engineering | 2010 | 27 | **3.5** |
| [S40] | Markus Klotzbücher, Nico Hochgeschwender, Luca Gherardi, Herman Bruyninckx, Gerhard Kraetzschmar, Davide Brugali. *The BRICS Component Model: A Model-based Development Paradigm for Complex Robotics Software Systems.* 28th Annual ACM Symposium on Applied Computing | 2013 | 09 | **2.5** |
| [S41] | John C. Georgas and Richard N. Taylor. *Policy-Based Self-Adaptive Architectures: A Feasibility Study in the Robotics Domain.* International Workshop on Software Engineering for Adaptive and Self-managing systems | 2008 | 35 | **3.0** |
| [S42] | Andreas Steck, Alex Lotz and Christian Schlegel. *Model-driven Engineering and Run-time Model-usage in Service Robotics.* 10th ACM International Conference on Generative Programming and Component Engineering | 2012 | 04 | **3.5** |
| [S43] | Francisco Ortiz, Diego Alonso, Bárbara Álvarez, Juan A. Pastor. *A Reference Control Architecture for Service Robots Implemented on a Climbing Vehicle.* 10th Ada-Europe International Conference on Reliable Software Technologies. | 2005 | 08 | **2.5** |
| [S44] | Jennifer Pérez, Nour Ali, Jose A. Carsı, Isidro Ramos, Bárbara Álvarez, Pedro Sanchez, Juan A. Pastor. *Integrating Aspects in Software Architectures: PRISMA Applied to Robotic tele-operated Systems.* Information and Software Technology. | 2007 | 31 | **3.5** |
| [S45] | Carle Côté, Dominic Létourneau, Clément Raïevsky, Yannick Brosseau, François Michaud. *Using MARIE for Mobile Robot Software Development and Integration.* Software Engineering for Experimental Robotics | 2007 | 12 | **3.5** |
| [S46] | Antonio C. Domínguez-Brito, Daniel Hernández-Sosa, José Isern-González, Jorge Cabrera-Gámez. *CoolBOT: A Component Model and Software Infrastructure for Robotics.* Software Engineering for Experimental Robotics | 2007 | 06 | **2.5** |

| [S47] | José Baca, Prithvi Pagala, Claudio Rossi, Manuel Ferre. **Modular Robot Systems Towards the Execution of Cooperative Tasks in Large Facilities**. Robotics and Autonomous Systems | 2015 | 2 | 3.5 |
|---|---|---|---|---|
| [S48] | Didier Crestani, Karen Godary-Dejean, Lionel Lapierre. **Enhancing Fault Tolerance of Autonomous Mobile Robots**. Robotics and Autonomous Systems | 2015 | 0 | 3.5 |
| [S49] | Andrea Bonarini, Matteo Matteucci, Martino Migliavacca, Davide Rizzi. **R2P: An Open Source Hardware and Software Modular Approach to Robot Prototyping.** Robotics and Autonomous Systems | 2014 | 06 | 4.0 |
| [S50] | Christian Berger. **From a Competition for Self-Driving Miniature Cars to a Standardized Experimental Platform: Concept, Models, Architecture, and Evaluation.** Journal of Software Engineering for Robotics | 2014 | 07 | 3.5 |
| [S51] | Jan Oliver Ringert, Alexander Roth, Bernhard Rumpe, Andreas Wortmann. **Code Generator Composition for Model-Driven Engineering of Robotics Component & Connector Systems.** International Workshop on Model-Driven Robot Software Engineering. | 2014 | 0 | 3.0 |
| [S52] | Peihua Chen, Qixin Cao. **A Middleware-Based Simulation and Control Framework for Mobile Service Robots.** Journal of Intelligent & Robotic Systems | 2014 | 0 | 3.5 |
| [S53] | Manja Lohse, Frederic Siepmann, Sven Wachsmuth. **A Modeling Framework for User-Driven Iterative Design of Autonomous Systems.** International Journal of Social Robotics | 2014 | 2 | 3.0 |
| [S54] | Gianluca Antonelli, Khelifa Baizid, Fabrizio Caccavale, Gerardo Giglio, Giuseppe Muscio, Francesco Pierri. **Control Software Architecture for Cooperative Multiple Unmanned Aerial Vehicle-Manipulator Systems.** Journal of Software Engineering for Robotics | 2014 | 0 | 3.5 |
| [S55] | Ion Mircea Diaconescu, Gerd Wagner. **Towards a General Framework for Modeling, Simulating and Building Sensor/Actuator Systems and Robots for the Web of Things**. International Workshop on Model-Driven Robot Software Engineering | 2014 | 0 | 2.5 |
| [S56] | L. Riazuelo, Javier Civera, J.M.M. Montiel. **$C^2$TAM: A Cloud Framework for Cooperative Tracking and Mapping.** Robotics and Autonomous Systems | 2014 | 18 | 4.0 |

# Appendix B

## Extended Details of Research Methodology

## B.1 Selection of Primary Studies

***Step I – Primary Search*** is further decomposed into four steps to identify the relevant literature. We provide a summary of the each step involved in the primary search process in Table I.

Table I. A Summary of Steps for Primary Search of Relevant Literature.

| Search Step | Description |
|---|---|
| *A.* **Deriving Search Terms** | We derived search terms from RQs in Section 3.1 |
| *B.* **Considering Synonyms and Alternatives for Search Terms** | We considered the alternative keywords for deriving the literature search stings:<br>- **Software Architecture** as [*Software Design*, *Software Component*, *Software Framework*]<br>The other relevant terms for architecture like 'Software Structure' or 'Software Styles' were excluded as they resulted in a large number of irrelevant literature hits.<br>- **Robots** as [Robotic, Robotics, Humanoid].<br>The other relevant terms for robots like 'Agent' or 'Android' were also excluded to avoid a large number of irrelevant hits. |
| *C.* **Combining Search Terms to Compose Search Strings** | We combined the search terms to compose the search strings:<br>- Boolean *OR* operators were used to incorporate alternative spellings and synonyms<br>- Boolean *AND* operators is used to link the search terms. |
| *D.* **Dividing and Customizing Search Strings** | We divided and customized the search strings so that they could be applied to different databases (digital libraries) containing the literature.<br>We assigned the unique IDs to every (sub-) search string. |

***Step II – Customized Search Strings***

As per the step D in Table I, we derived the following customized search strings as per the individual digital libraries. Google Scholar have been used as a primary source for the identification of relevant literature in systematic reviews. However, among others the critical factors such as the (i) frequency of changes (tweaks) to Google search algorithm[9], (ii) personalized and context-driven searching capabilities, and (iii) a significant amount of gray literature determined our decision about not including Google Scholar as one of the primary literature search source. We only used Google Scholar as an auxiliary search engine to ensure if some relevant literature may not have been missed by our selected search engines/digital libraries (cf. Figure 3) – only randomly cross-checking the search results. The cross check did not reveal any additional relevant study.

- ***IEEE eXplore*** (www.ieeexplore.ieee.org)

Search String 1 for IEEE eXplore

("Document Title": Software *OR* "Abstract": Software) *AND* ("Document Title": Architecture *OR* "Document Title": Component *OR* "Document Title": Design *OR* "Document Title": Framework *OR* "Abstract": Architecture *OR* "Abstract": Component *OR* "Abstract": Design *OR* "Abstract": Framework) *AND* ("Document Title": Robot *OR* "Document Title": Robotic *OR* "Document Title": Humanoid *OR* "Abstract": Robot *OR* "Abstract": Robotic *OR* "Abstract": Humanoid)

- ***ACM Digital Library*** (www.dl.acm.org)

---

[9] Google Algorithm Change History: http://www.seomoz.org/google-algorithm-change

Search String 2 for ACM Digital Library

((Owner:ACM) **AND**(Abstract "Software" **OR** Title "Software") **AND** (Abstract "Architecture" **OR** Abstract "Component" **OR** Abstract "Design" **OR** Abstract "Framework" **OR** Title "Architecture" **OR** Title "Component" **OR** Title "Design" **OR** Title "Framework") **AND** (Abstract "Robot" **OR** Abstract "Robotic" **OR** Abstract "Humanoid" **OR** Title "Robot" **OR** Title "Robotic" **OR** Title "Humanoid"))

- ***Springer Link*** (www.link.springer.com)

Search String 3 for Springer Link

(Software) **AND** (Architecture **OR** Design **OR** Component **OR** Framework) **AND** (Robot **OR** Robotic **OR** Humanoid)

- ***Science Direct*** (www.sciencedirect.com)

Search String 4 for Science Direct

TITLE-ABSTR-KEY((Software) **AND** (Architecture **OR** Design **OR** Component **OR** Framework) **AND** (Robot **OR** Robotic **OR** Humanoid))

- ***Scopus*** (www.scopus.com)

We have refined the following search string with the exclusion of the term 'KEY' (keyword-based selection), that helped us eliminating a significant number of irrelevant studies. In doing so, we may have overlooked some relevant literature (cf. Section 9, *Validity Threats to the Identification of Primary Studies*). However, we believe the refinement of the search strings has helped us to minimize the irrelevant literature.

Search String 5 for Scopus

TITLE-ABS((Software) **AND** (Architecture **OR** Design **OR** Component **OR** Framework) **AND** (Robot **OR** Robotic **OR** Humanoid))

***Step III – Secondary Search*** a two phase process consisting of the *primary* and *secondary* search is based on the guidelines and empirical comparison of literature search using *digital database*/libraries and the *snowballing process* [45]. As detailed earlier, the primary search (cf. Section 3.2, Figure 3) identified the relevant studies for SMS by executing the search strings on digital databases/libraries. After screening and qualitative assessment, we selected 56 studies for review and analyzed the references/bibliography section for each of the 56 selected studies to find other relevant studies for their possible inclusion – the snowballing process. We limited the snowballing process to the selected primary (56) studies to avoid any exhaustive search. A detailed comparison and the relative benefits and limitations of literature search using digital databases vs snowballing is detailed in [45]. As a results of snowballing process, we did not identify any study to be included in the review. In particular, during the snowballing process; we found studies that were either (i) already included in the list of primary studies, (ii) not explicitly relevant to RQs or (iii) eliminated during the qualitative assessment process.

## B.2 Screening and Qualitative Assessment of Studies

The study selection comprises of a two-step process that includes screening and qualitative assessment as presented in Tables II and Table III. In Table III, the qualitative assessment helps us to include/exclude studies and rank the selected studies based on their quality score in the Appendix A.

**Step I – Screening of Studies**

The study screening is a two-step process consisting of the *Generic Screening* (having five sub-steps, **I-A** to **I-E**) and *Specific Screening* as outlined in Table II. During the Generic Screening step (review of study titles), first we need to screen the 308 studies to ensure that (i) *duplicate studies* (Step I-A), (ii) *non English language* literature (Step I-B), (iii) *non peer-reviewed* and *non-published research* (Step-I-C) and, (iv) any study representing an *entire book* (Step I-D) are removed first. Furthermore, any *secondary study* (such as a systematic review or survey-based literature) is eliminated from primary studies to be reviewed – any secondary study represents related research to our SMS (detailed in Section 2.3). The Generic Screening step helped us to remove a significant number of studies (211) with total remaining studies 97.

During the Step II - Specific Screening, we performed a preliminary review to analyze the relevance of the studies to RQs (method, technique or a solution for robotic software systems). During Step II, the decision to exclude ([NO]) or proceeding to the final selection ([YES]) was based on an examination of study *titles* and a preliminary review of the *abstracts*, *conclusions* and any other relevant part of the remaining studies. Based on the screening the number of studies was reduced to 97.

Table II. Summary of the Study Selection Process (without qualitative assessment)

| | Step I – Generic Screening | | |
|---|---|---|---|
| I-A | Is the study a duplicate? | YES | NO |
| I-B | Is the study in English language? | YES | NO |
| I-C | Is the study a scientific peer-reviewed published research (no white papers or technical reports)? | YES | NO |
| I-D | Is the study not a secondary study? | YES | NO |
| I-E | Is the study not a book? | YES | NO |
| | If [YES] to all four criteria then go to Step II, otherwise exclude study | | |
| | Step II – Specific Screening | | |
| II-A | RQ1, RQ2 *Does the study presents an architectural method, technique or a solution for robotic software systems?* | | |
| | If [**YES**] go to Table B.2 (qualitative assessment), otherwise exclude study | | |

### Step II – Qualitative Assessment of Studies

During qualitative assessment of 97 included studies, we focused on assessing the technical rigor of contents presented in the study. The qualitative assessment is based on two factors as general assessment (G) and specific assessment (S), in Table III. Quality scores provided us with a numerical quantification to rank the selected studies in the Appendix. We adopted the guidelines of qualitative synthesis of research evidence [50], tailored them as per the needs for our study as presented in Table III.

Quality assessment represents 5 factors criteria, providing a maximum score of 1. In the assessment formula below, S and G each represent a total of five factors as **Specific and Generic Items** with S having a maximum score of 3 and G with a maximum score 1. S contributes three times more than G (75% weight) as specific contributions of a study are more important than general factors for assessment. Based on the consensus among the researchers the maximum score was decided as G + S = 4, where a 3 – 4 score represented quality papers, a score less than 3 and greater than or equal to 1.5 was acceptable and a score less than 1.5 resulted in study exclusion.

$$Quality\ Score = \left[ \frac{\sum_{G=1}^{5}}{5} + \left( \frac{\sum_{S=1}^{5}}{5} \times 3 \right) \right]$$

Based on qualitative assessment of *97* studies, we excluded a total of *41* (quality score less than 1.5) studies to finally select *56* primary studies for the review.

Table III. Summary of Quality Assessment Checklist

| General Items for Quality Assessment (G) | | | | |
|---|---|---|---|---|
| | Score for General Items | Yes = 1 | Partial = 0.5 | No = 0 |
| G1 | Are *problem definition* and *motivation* of the study clearly presented? | | | |
| G2 | Is the *research environment* in which the study was carried out properly explained? | | | |
| G3 | Are *research methodology* and its *organization* clearly stated? | | | |
| G4 | Are the *contributions* of the in-line with presented *results*? | | | |
| G5 | Are the *insights* and *lessons learnt* from the study explicitly mentioned? | | | |
| Specific Items for Quality Assessment (S) | | | | |
| | Score for Specific Items | Yes = 1 | Partial = 0.5 | No = 0 |
| S1 | Is the research clearly *focused on software architecture solutions for robotic systems*? | | | |
| S2 | Are the details about *related research* clearly addressing *architectural solutions*? | | | |
| S3 | Is the *research validation* clearly illustrates the evaluation of architectural solutions? | | | |
| S4 | Are the results *clearly validated* in a real (industrial case study) evaluation context? | | | |
| S5 | Are limitations and future research clearly positioned? | | | |

## B.3 Data Extraction for Synthesis

To collect and record the format in Table IV collects two types of data (i) *generic and study demographic data items* (D01 – D06) and ii) *classification and mapping specific data items* (M01 – M10). The latter category helps us to answer the RQs (cf. Section 3.1).

Table IV. Template for Extraction of Data from Primary Studies

| ID | Data Item | Aim |
|---|---|---|
| Generic and Study Demographic Data Items | | |
| D01 | *Study ID* | Unique id of study ___________ |
| D02 | *Study Title* | The title of Study ___________ |
| D03 | *Bibliography* | a) List of Author(s) ___________<br>b) Year of Publication ___________<br>c) Source of Publication<br>*Journal* □ *Conference* □ *Symposium or Workshop* □ *Other* _______ |
| D04 | *Citation Count* | Total number of citations ___________ |
| D05 | *Quality Score* | Quality score of study ___________ |
| D06 | *Additional Information* | Any additional or study specific information ___________ |
| Classification and Mapping Specific Data Items | | |
| M01 | *Research Problem* | Overview of research problems addressed ___________ |
| M02 | *Architectural Solution* | Overview of solution to address the problem ___________ |
| M03 | *Research Context* | Context and application domain:<br>*Academic* □ *Industrial* □ *Both* □ *Other* □ |
| M04 | *Framework Support* | Architectural Framework for Development.<br>*Yes* ○ *No* ○ *Description* ___________ |
| M05 | *Modeling Notations* | *UML* □ *ADL* □ *Graph Models* □ *Ontologies* □ *Other* |
| M06 | *Architecture Model Type* | *Component* □ *Service* □ *Object* □ *Other* _______ |
| M07 | *Validation Method* | *Design and Evaluation* □ *Case Study* □ *Survey* □ *Experiments* □ *Other* _______ |
| M08 | *Architecture Evaluations* | *Yes* ○ *No* ○ *Description* ___________ |
| M09 | *Research Trends* | The identified research trends ___________ |
| M10 | *Future Dimensions* | The identified future research ___________ |